\documentstyle[preprint,aps,prb]{revtex}
\begin{document}
\draft
\tightenlines 
\title{Resonant Raman scattering off neutral quantum dots}
\author{Alain Delgado\cite{alain}$^1$, Augusto Gonzalez
 \cite{augusto}$^2$, and Eduardo Men\'endez-Proupin\cite{eduardo}$^3$}
\address{$^1$Centro de Estudios Aplicados al Desarrollo Nuclear,
 Ciudad Habana, Cuba}
\address{$^2$Instituto de Cibernetica, Matematica y Fisica, Calle E 
 309, Vedado, Habana 4, Cuba}
\address{$^3$Instituto de Materiales y Reactivos, 
 Universidad de La Habana, Cuba}
\date{Received: \today}

\maketitle

\begin{abstract}
Resonant inelastic (Raman) light scattering off
neutral GaAs quantum dots which contain a mean number, $N=42$, of 
electron-hole pairs is computed. 
We find Raman amplitudes corresponding to
strongly collective final states (charge-density excitations) of
similar magnitude as the amplitudes related to 
weakly collective or single-particle excitations. 
As a function of the incident laser frequency or the magnetic field, 
they are rapidly varying amplitudes.
It is argued that strong Raman peaks should come out in the 
spin-density channels, not related to valence-band mixing effects
in the intermediate states.
\end{abstract}
\vspace{.5cm}

\section{Introduction}

The experimental study of the inelastic (Raman) scattering of light
in arrays of quantum dots (qdots) began a few years ago 
\cite{Lockwood,Schuller}. These studies were aimed at investigating 
multipole excitations or spin excited states in the quantum dot, which
do not leave traces in luminescence or absorption experiments. In both
papers, quasi-bidimensional dots with radii around 100 nm and nominal
electron density about $8\times 10^{11}~{\rm cm}^{-2}$ were 
studied. A rich spectrum of single-particle (SPE), and collective
charge-density (CDE) and spin-density excitations (SDE) was observed.

The single-particle spectra were interpreted in terms of the 
single-particle density of states in the dot \cite{Lockwood}, computed
in the Hartree-Fock (HF) approximation. The energy position and Raman
strengths of CDE states in dots with 12 electrons, computed within 
time-dependent HF theory \cite{Steinebach}, were shown to qualitatively 
agree with the experimental results. More consistent calculations in the 
framework of time-dependent density-functional theory (DFT) were carried
out in Ref. \onlinecite{Emperador}, where the charge and spin dynamic structure
functions were computed for a system of 200 electrons. The multipolarities 
of the observed CDE and SDE peaks, and the relative peak intensities, 
as functions of the transferred wave vector of the light, were reproduced.

Let us stress that, in electron qdots, Raman processes in which the 
final states are SDE require the account for valence-band mixing in the
intermediate hole state. For this reason, CDE peak intensities reported
in Ref. \onlinecite{Emperador} can not be properly compared with intensities in the
SDE channels. The inclusion of valence-band mixing effects in the 
intermediate states of Raman processes is to be published elsewhere
\cite{Alain}.

In the present paper, we compute Raman amplitudes for neutral quantum
dots, where the number of holes, $N$, equals the number of electrons 
in the dot. To the best of our knowledge, there are no similar 
calculations in the literature.

Although the lifetime of the excitons may pose certain difficulties to
the Raman measurements, we believe that it does not represent a real
challenge to present experiments. With relative independence on the
dot parameters, the multi-excitonic system reaches typical densities
around $5\times 10^{11}$ pairs/cm$^2$, a value which may be obtained by
pumping a neutral qdot with a stationary mean-power laser. In fact,
densities well above $10^{12}~{\rm cm}^{-2}$ have been achieved already for a 
few years \cite{Wolfe} with pulsed high-power lasers.

Due to the lack of experimental results, we will focus on the qualitative
aspects following from our calculations. A simplified two-band model
of a disk-shaped qdot with parabolic confinement \cite{WHFJ96} is to be 
used.

We shall, first, make some remarks concerning the computational method.
The Random Phase Approximation (RPA) is the common frame, as in the 
cited papers. We adopt the wave-function approach of Nuclear Physics 
\cite{Ring-Schuck}, and construct RPA approximations to the wave
functions of both final and intermediate states. Coulomb interactions and 
collective effects are exactly accounted for within the RPA, even for the
intermediate states entering the Raman amplitudes, which are states with 
$N+1$ e-h pairs. Corrections to the RPA functions, such as the e-h
pairing correlations, could, in principle, be included by means of the
quasiparticle RPA scheme \cite{Ring-Schuck}. 
The electron-radiation (e-r) interaction hamiltonian is written in second 
quantization in the basis of HF single-particle states. Our treatment
of Raman scattering follows the lines of Refs. \onlinecite{Trallero} 
and \onlinecite{Eduardo}, in the sense that the e-r interaction
causes transitions between multi-excitonic states. Thus, energy 
denominators contain $N$-pair instead of single-particle energies.

Concerning the numerical results, there are a few points to stress. 
First, the absorption threshold, or the frequency for which extreme
resonance is achieved in Raman scattering, grows at a rate of 0.3 meV 
per pair added to the dot. This is an indirect way of determining the 
mean number of pairs in the dot. Second, Raman peaks in quadrupole 
channels are 1/10 of monopole peaks at momentum transfer around 
$0.8\times 10^5 ~{\rm cm}^{-1}$. Next, intensities corresponding to 
weakly collective or SPE are comparable in magnitude to the strongest
CDE peaks, and vary very rapidly with the magnetic field or the frequency
of the incident laser. Thus, our calculated spectra resemble more the
complex spectra of quantum wells in strong magnetic fields \cite{Davies}, 
and differ from the smooth experimental curves obtained in 
Ref. \onlinecite{Schuller} for the pure electronic strong-confinement
qdot.

A last point which deserves attention is the fact that Raman scattering
in SDE channels do not require mixing of hole bands. Thus SDE peaks 
should be observed in any polarization of the scattered light.

The present calculation of Raman cross sections completes a series of 
papers on the optical properties of the $N$-pair system in a qdot. The 
dominance of a giant dipole resonance in the infrared absorption,
which position scales as $N^{1/4}$, was shown in Ref. \onlinecite{gdr}.
This resonance could be studied through the modulations of the 
photoluminescence caused by an infrared source, in the same way as the
infrared excitations in the ``$N$-electron plus one hole'' system are
presently studied \cite{Nickel}. On the other hand, the position and
intensity of the coherent magneto-luminescence peak were computed in
Ref. \onlinecite{Boris y yo}.

The plan of the paper is as follows. The basic expressions for Raman
cross sections along with HF, RPA and particle-particle Tamm-Dancoff
approximations (pp-TDA) are presented in Section II. The formalism is
well established in the Nuclear Physics context \cite{Ring-Schuck}.
We underline in that section only the main points for the sake of 
completeness. Computed ground-state properties, the multipole excitations
and their strength variations with magnetic field, the band gap
renormalization as a function of the number of pairs, and the Raman
cross sections are given in Section III. Final remarks are presented 
at the end.

\section{The basics}

The resonant inelastic (Raman) light scattering off a
neutral quantum dot containing a mean number, $N$, of 
electron-hole pairs is schematically represented in Fig. \ref{fig1}. The
energy of the incident photon, $\hbar\omega_i$, is supposed to be close 
to the band gap energy, $E_{gap}$.
In Fig. \ref{fig1} (a), the final state has the same spin quantum numbers 
as the initial (ground) state of the $N$-pair system. It is, in general, a 
charge-density excitation (CDE). The electron and hole spins are 
represented by arrows. Incident and emitted photons are drawn as wavy 
lines. Additionally, there are also Raman processes in which the final states 
involve changes in the spin quantum numbers. These states will be called
spin-density excitations (SDE). We have represented in Fig. \ref{fig1} (b)
a situation in which the total electron and total hole spin projections
experience changes $\Delta S_{ze}=-1$ and $\Delta S_{zh}=+1$, respectively.

The amplitude for the Raman processes depicted in Fig. \ref{fig1} (a) is given 
by:

\begin{equation}
A_{fi}^{CDE}=\sum_{int} \frac{\sqrt{N_f+1}~_N\langle f|H^+_{e-r}| int
 \rangle_{N+1}~_{N+1}\langle int|H^-_{e-r}|i\rangle_N \sqrt{N_i}}
 {\hbar\omega_i-(E_{int}-E_i)+i\Gamma},
\label{A-CDE}
\end{equation}

\noindent
where $N_i$, $N_f$ are the mean number of photons in the initial and
final states. For spontaneous Raman scattering, $N_f=0$. The sum runs over
intermediate states 
with $N+1$ pairs and the appropriate quantum numbers. $\Gamma$ is the  
lifetime broadening. We will take it phenomenologically
as $\Gamma=0.5$ meV. \cite{Steinebach} The resonance condition means that the 
leading contribution to (\ref{A-CDE}) comes from intermediate states
satisfying

\begin{equation}
\hbar\omega_i\approx E_{int}-E_i,
\end{equation}

\noindent
i. e. the incident photon has nearly the same energy as the jump in energy
from the initial to the intermediate quantum-dot state. 

$H_{e-r}^-$ is the interaction hamiltonian corresponding to the 
annihilation of a photon and creation of a new electron-hole pair. Its 
matrix element is written as

\begin{eqnarray}
_{N+1}\langle int|H^-_{e-r}|i\rangle_N = \frac{e}{m_0}
 \sqrt{\frac{2\pi\hbar}{V\omega_i\eta_i^2}}~\sum_{\alpha,\gamma} 
 \left(\vec\varepsilon_i\cdot\vec p_{\alpha\gamma}\right)
 \left(\int e^{i\vec q_i \cdot\vec r} \phi_{\alpha e}^*(\vec r) 
 \phi_{\gamma h}^*(\vec r) {\rm d}^3r\right)\nonumber\\ 
\times _{N+1}\langle int|e^{\dagger}_{\alpha} h^{\dagger}_{\gamma}
 |i\rangle_N.
\label{Her-}
\end{eqnarray}

In the normalization factor entering Eq. (\ref{Her-}), $e$ is the
electron charge, $m_0$ is the electron mass in vacuum, $V$ is the sample 
volume, and $\eta_i$ is the refraction index at frequency $\omega_i$. 
On the other hand, the first
factor in the sum comes from the (band) spin quantum numbers of the initial 
and final states in the interband transition. We give its detailed 
expression in Appendix \ref{Appendix A}. $\vec \varepsilon_i$ is the 
polarization vector of the incident light.

The next factor in the sum depends on the orbital (envelope) one-particle 
wave functions. As a basis for the one-particle states, we will use the
Hartree-Fock (HF) set, $\phi_{\alpha e}$ and $\phi_{\gamma h}$ for 
electrons and holes respectively. $\vec q_i$ is the wave vector
of the incident light. Detailed expressions for the orbital factor are 
given also in Appendix \ref{Appendix A}.

Finally, we will compute the last factor in (\ref{Her-}) by means of the
so called particle-particle Tamm-Dancoff Approximation (pp-TDA) formalism
\cite{Ring-Schuck}, to be briefly described below.

The second matrix element entering Eq. (\ref{A-CDE}) is written as

\begin{eqnarray}
_N\langle f|H^+_{e-r}|int\rangle_{N+1} = \frac{e}{m_0}
 \sqrt{\frac{2\pi\hbar}{V\omega_f\eta_f^2}}~\sum_{\alpha,\gamma} 
 \left(\vec\varepsilon_f^{~*}\cdot\vec p_{\alpha\gamma}^{~*}\right)
 \left(\int e^{-i\vec q_f \cdot\vec r} \phi_{\alpha e}(\vec r) 
 \phi_{\gamma h}(\vec r) {\rm d}^3r\right)\nonumber\\ 
 \times _N\langle f|h_{\gamma}e_{\alpha}|int\rangle_{N+1}.
\label{Her+}
\end{eqnarray}

\noindent
Its computation involves similar band and orbital factors (see Appendix 
\ref{Appendix A}).
The last factor requires -- besides the intermediate $(N+1)$-pair states,
to be obtained from the pp-TDA equations -- the knowledge of the $N$-pair
excited states, which will be obtained from the ordinary or ``particle-hole''
RPA formalism \cite{Ring-Schuck}. The starting point for both RPA and 
pp-TDA schemes are the HF single-particle states.

\subsection{The HF equations}

We take for the HF single-particle functions the following ansatz:

\begin{equation}
\phi_{\alpha}^{e(h)}=\sqrt{\frac{2}{L}} \sin \left(\frac{\pi z}{L}\right)
 \sum_s C_{\alpha,s}^{e(h)}~\chi_s (\vec r_{\Vert }),
\label{expansion}
\end{equation}

\noindent
where $0\le z\le L$, $L=12$ nm is the height of our disk-shaped qdot, 
$\vec r_{\Vert}$ is the projection of $\vec r$ onto the $xy$ plane, and
$\chi_s$ are the two-dimensional (2D) oscillator wave functions, given
elsewhere \cite{gdr}. The $C_{\alpha,s}$ coefficients are obtained from
the equations \cite{gdr,tesis Alain}:

\begin{eqnarray}
\sum_t &&\left\{ E_{es}^{(0)}\delta_{st} + \beta\sum_{\gamma\le \mu_F^e} 
 \sum_{u,v} \left[ \langle s,u|1/r|t,v\rangle-\langle s,u|1/r|v,t\rangle
 \right]C_{\gamma,u}^e C_{\gamma,v}^e \right. \nonumber\\
 &-& \left. \beta \sum_{\gamma\le\mu_F^h}\sum_{u,v} 
 \langle s,u|1/r|t,v\rangle C_{\gamma,u}^h C_{\gamma,v}^h\right\}
 C_{\alpha,t}^e=E_{e \alpha} C_{\alpha,s}^e,
\label{HF}
\end{eqnarray}

\noindent
and a similar set of equations for the $C_{\alpha,s}^h$. Notice that the
$E^{(0)}$ are 2D oscillator energies:

\begin{eqnarray}
E_{es}^{(0)}&=&\hbar\sqrt{\omega_0^2+\omega_c^2/4}~\{ 2 k_s+|l_s|+1\}+
 \frac{\hbar\omega_c}{2} l_s+g_e\mu_B B S_z^e,\\
E_{hs}^{(0)}&=&\frac{m_e}{m_h}~\hbar\sqrt{\omega_0^2+\omega_c^2/4}~
 \{ 2 k_s+|l_s|+1\}-\frac{m_e}{m_h}\frac{\hbar\omega_c}{2} l_s
 -g_h\mu_B B S_z^h,
\end{eqnarray}

\noindent
where $\hbar\omega_0=3$ meV is the in-plane confinement, 
$\omega_c=eB/(m_e c)$
is the electron cyclotronic frequency, and $g_e,~g_h$ are Lande 
factors. We took parameters appropriate for GaAs: $m_e=0.067 m_0$, i. e.
$\hbar\omega_c/B=1.728$ meV/Teslas, the ratio of in-plane masses is 
$m_e/m_h=0.067/0.11$, and $g_e\mu_B=-0.0173$ meV/Teslas, 
$g_h\mu_B=0.0296$ meV/Teslas.
In the $g$-factors, the effect of qdot height was approximately accounted
for \cite{Boris}.

On the other hand, the $\langle s,u|1/r|t,v\rangle$ are Coulomb matrix 
elements taken over 2D oscillator wave functions \cite{gdr}, and the 
strength $\beta$ is given by $0.8~e^2/(\kappa l_0)$, where:

\begin{equation}
l_0=\sqrt{ \frac{\hbar}{ m_e \sqrt{\omega_0^2+\omega_c^2/4} } },
\label{length}
\end{equation}

\noindent
is the unit of length, $\kappa=12.5$ is the dielectric constant, 
and the 0.8 coefficient takes care approximately of the effect on 
Coulomb interaction of averaging over the z coordinate \cite{MacD}.

Equations (\ref{HF}) are solved iteratively. We start by occupying the 
lowest oscillator shells, construct the matrix inside brackets in 
(\ref{HF}), and
iterate until convergence is reached. The occupation of HF levels is
actualized after every 10 steps in accordance to the current values of the
HF energies. 15 oscillator shells are used in the calculations, i. e. a
total of 240 2D oscillator states.

\subsection{The RPA equations}
\label{RPA}

In the RPA \cite{Ring-Schuck}, we allow a general correlated 
ground state, $|RPA\rangle$, and the excited states are looked
for in the form

\begin{equation}
\Psi = Q^{\dagger} |RPA\rangle ,
\end{equation}

\noindent
where the $Q^{\dagger}$ operator for CDE states is given by the 
expression

\begin{equation}
Q^{\dagger}_{CDE}=\sum_{\sigma,\lambda}(X^e_{\sigma\lambda} 
 e^{\dagger}_{\sigma} e_{\lambda}+X^h_{\sigma\lambda}
 h^{\dagger}_{\sigma} h_{\lambda}-Y^e_{\lambda\sigma} 
 e^{\dagger}_{\lambda} e_{\sigma}-Y^h_{\lambda\sigma} 
 h^{\dagger}_{\lambda} h_{\sigma}).
\label{Q}
\end{equation}

\noindent
The index $\lambda$ runs over occupied HF states, 
and $\sigma$ runs over unoccupied states. The $X,~Y$ coefficients
are nonzero only for transitions respecting the selection rules, i.e. the
spin projection doesn't change, and the change in angular momentum is
fixed (a given multipolarity of the excitation). These coefficients 
satisfy the equations:

\begin{eqnarray}
\sum_{\tau,\mu}\left\{ A^{ee}_{\sigma\lambda,\tau\mu} X^e_{\tau\mu}
 +A^{eh}_{\sigma\lambda,\tau\mu} X^h_{\tau\mu}
 +B^{ee}_{\sigma\lambda,\mu\tau} Y^e_{\mu\tau}
 +B^{eh}_{\sigma\lambda,\mu\tau} Y^h_{\mu\tau}\right\}&=&
 \hbar\Omega X^e_{\sigma\lambda},\nonumber\\
 \sum_{\tau,\mu}\left\{ A^{he}_{\sigma\lambda,\tau\mu} X^e_{\tau\mu}
 +A^{hh}_{\sigma\lambda,\tau\mu} X^h_{\tau\mu}
 +B^{he}_{\sigma\lambda,\mu\tau} Y^e_{\mu\tau}
 +B^{hh}_{\sigma\lambda,\mu\tau} Y^h_{\mu\tau}\right\}&=&
 \hbar\Omega X^h_{\sigma\lambda},\nonumber\\
 \sum_{\tau,\mu}\left\{ B^{ee}_{\lambda\sigma,\tau\mu} X^e_{\tau\mu}
 +B^{eh}_{\lambda\sigma,\tau\mu} X^h_{\tau\mu}
 +A^{ee}_{\lambda\sigma,\mu\tau} Y^e_{\mu\tau}
 +A^{eh}_{\lambda\sigma,\mu\tau} Y^h_{\mu\tau}\right\}&=&
 -\hbar\Omega Y^e_{\lambda\sigma},\nonumber\\
 \sum_{\tau,\mu}\left\{ B^{he}_{\lambda\sigma,\tau\mu} X^e_{\tau\mu}
 +B^{hh}_{\lambda\sigma,\tau\mu} X^h_{\tau\mu}
 +A^{he}_{\lambda\sigma,\mu\tau} Y^e_{\mu\tau}
 +A^{hh}_{\lambda\sigma,\mu\tau} Y^h_{\mu\tau}\right\}&=&
 -\hbar\Omega Y^h_{\lambda\sigma},
\label{RPAeq}
\end{eqnarray}

\noindent
in which $\hbar\Omega$ is the excitation energy,  $\tau$ and $\mu$ are 
indexes similar to $\sigma$ and $\lambda$, respectively, and the $A$ 
and $B$ matrices are given by \cite{gdr,tesis Alain}:

\begin{eqnarray}
A^{ee}_{\sigma\lambda,\tau\mu} &=& (E_{e\sigma}-
 E_{e\lambda}) \delta_{\sigma\tau} \delta_{\lambda\mu}
 + \beta \left( \langle \sigma,\mu |1/r|\lambda,\tau\rangle -\langle 
 \sigma,\mu |1/r|\tau,\lambda\rangle \right),\nonumber\\
A^{eh}_{\sigma\lambda,\tau\mu} &=& -\beta\langle \sigma,\mu |
 1/r|\lambda,\tau \rangle,\nonumber\\
B^{ee}_{\sigma\lambda,\mu\tau} &=& \beta ( \langle \sigma,\tau
|1/r|\lambda,\mu\rangle -\langle \sigma,\tau |1/r|
\mu,\lambda\rangle ),\nonumber\\
B^{eh}_{\sigma\lambda,\mu\tau} &=& -\beta\langle \sigma,\tau 
|1/r|\lambda,\mu \rangle. 
\end{eqnarray}

\noindent
Notice, for example, that in $A^{eh}_{\sigma\lambda,\tau\mu}$, $\sigma$
and $\lambda$ are electronic HF states, and $\tau$, $\mu$ -- hole states.
$A^{hh}$ has formally the same expression as $A^{ee}$, $A^{he}$ the same 
as $A^{eh}$, etc. Let us stress also that Coulomb matrix elements over
HF states enter the RPA equations (\ref{RPAeq}), they can be computed 
from the matrix elements over oscillator states by means of the 
expansions (\ref{expansion}).
Usually, positive (physical) and negative (unphysical) excitation energies 
come from (\ref{RPAeq}). The physical solutions annihilate the
RPA ground state

\begin{equation}
Q |RPA\rangle = 0,
\end{equation}

\noindent
and satisfy the normalization condition

\begin{equation}
1= \sum_{\sigma,\lambda} 
 \{ |X_{\sigma\lambda}^e|^2+|X_{\sigma\lambda}^h|^2
 -|Y_{\lambda\sigma}^e|^2-|Y_{\lambda\sigma}^h|^2 \} .
\end{equation}
 
To evaluate the collective character of a state $\Psi$, we compute the 
matrix elements of the multipole operators: $\langle\Psi|D_l|RPA\rangle$.
Collective states give significant transition strengths, whereas 
single-particle excitations give practically zero matrix elements. The
multipole operator, $D_l$, is defined as:

\begin{equation}
D_l=e \sum_{\alpha,\gamma}\left\{ d^{hl}_{\alpha\gamma} 
 h_{\alpha}^{\dagger} h_{\gamma}-d^{el}_{\alpha\gamma}
e_{\alpha}^{\dagger} e_{\gamma} \right\},
\end{equation}

\noindent
where 

\begin{eqnarray}
d^l_{\alpha\gamma}&=&\langle \alpha|\rho^{|l|} e^{i l \theta}
 |\gamma\rangle;~~l\ne 0,\nonumber \\
&=&\langle \alpha|\rho^2|\gamma\rangle;~~l=0.
\end{eqnarray}

\noindent
$\rho$ and $\theta$ are polar coordinates in the $xy$ plane. Detailed 
expressions for $d^l_{\alpha\gamma}$ are given in Appendix 
\ref{Appendix B}. 

Multipole matrix elements are computed from the RPA amplitudes in the
following way:

\begin{equation}
\langle \Psi |D_l| RPA\rangle = e \sum_{\sigma,\lambda} 
 \left\{ X^{h*}_{\sigma\lambda} d^{hl}_{\sigma\lambda}-
 X^{e*}_{\sigma\lambda} d^{el}_{\sigma\lambda}+
 Y^{h*}_{\lambda\sigma}d^{hl}_{\lambda\sigma}-
 Y^{e*}_{\lambda\sigma}d^{el}_{\lambda\sigma} \right\}.
\end{equation}

They fulfill the energy-weighted sum rules \cite{Ring-Schuck}

\begin{eqnarray}
\sum_{\Psi}\hbar\Omega_{\Psi}~\left\{ |\langle \Psi |D_l|RPA \rangle|^2
 +|\langle \Psi |D_{-l}|RPA \rangle|^2\right\}&=&2 \hbar^2 e^2 l^2
 \left\{ \frac{1}{m_e}\sum_{\lambda\le \mu_F^e}
 \langle\lambda|(r^2)^{|l|-1}|\lambda\rangle\right.\nonumber\\
 &+&\left. \frac{1}{m_h}\sum_{\lambda\le \mu_F^h}
 \langle\lambda|(r^2)^{|l|-1}|\lambda\rangle\right\},
\label{S1}
\end{eqnarray}

\noindent
for $l\ne 0$, and

\begin{equation}
\sum_{\Psi}\hbar\Omega_{\Psi}~ |\langle \Psi |D_0|RPA \rangle|^2
 =2 \hbar^2 e^2 \left\{ \sum_{\lambda\le \mu_F^e}
 \langle\lambda|\frac{r^2}{m_e}|\lambda\rangle+
 \sum_{\lambda\le \mu_F^h}\langle\lambda|\frac{r^2}{m_h}|\lambda\rangle
 \right\},
\label{S10}
\end{equation}

\noindent
for $l=0$. The $\mu_F$'s are Fermi levels. Thus, $\lambda\le \mu_F$
means that the sum runs over occupied HF states.
Explicit evaluation of the r. h. s. of (\ref{S1},\ref{S10}) 
is done in Appendix \ref{Appendix B}.

Spin excitations can also be built on within the RPA formalism. For
example, a state with $\Delta S_{ez}=1$, $\Delta S_{hz}=0$  can be 
obtained from a $Q^{\dagger}$ like (\ref{Q}) with only electron operators,
such that the transitions satisfy the spin selection rule. It should be
noticed, however, that the simple combination of one-particle excitations 
in Eq. (\ref{Q}) does not allow us to construct ``$2p-2h$'' excited states
with $\Delta S_{ez}=1$, $\Delta S_{hz}=-1$, for example, entering the
final states of Raman SDE processes. 

\subsection{pp-TDA}

The pp-TDA scheme allows us to build up states with $2N+2$ particles
starting from the the ground state of the $N$-pair system \cite{Ring-Schuck}. 
Notice that there are 12 possibilities 
for the added pair of particles. We can add, for example, an 
e-e pair with various spin orientations. In this subsection, we focus on 
the situations where an optically created e-h pair is added. That is, only
the following two possibilities are considered: e$\uparrow$h$\downarrow$ or
e$\downarrow$h$\uparrow$.

The $Q^{\dagger}$ operator, analogous to Eq. (\ref{Q}), is written in 
the following form:

\begin{equation}
Q^{\dagger}=\sum_{\sigma,\tau} V^{(N+1)}_{\sigma\tau} e^{\dagger}_{\sigma}
 h^{\dagger}_{\tau},
\end{equation}

\noindent
where, as before, $\sigma$ and $\tau$ label states above the Fermi levels.
$Q^{\dagger}$ acts on the RPA ground state to produce states with $N+1$ pairs.
The $V$ coefficients satisfy the equations:

\begin{eqnarray}
(\hbar\Omega-E^e_{\sigma}-E^h_{\tau}) V^{(N+1)}_{\sigma\tau}&=&
 -\beta\sum_{\sigma',\tau'} \langle \sigma,\tau|1/r|\sigma',\tau'\rangle
 V^{(N+1)}_{\sigma'\tau'}.
\label{pp-RPA}
\end{eqnarray}

\noindent
The quantity $\hbar\Omega$ gives the excitation energy, measured with
respect to the $|RPA\rangle$ $N$-pair state: $E(N+1)-E_{RPA}(N)$. 

RPA and pp-TDA excitations energies and coefficients $X$, and $V$ are to
be used in the computation of Raman amplitudes.

\subsection{Raman scattering in CDE channels}

The inelastic scattering of light, schematically represented in Fig. 
\ref{fig1} (a), is characterized by a Raman shift: 
$\hbar\omega_i-\hbar\omega_f=E_f-E_i=\hbar\Omega_f$. 
The amplitude for the process is given by Eq. (\ref{A-CDE}).
This amplitude will depend on the scattering angles.

To state a convention, the dot plane will define the $xy$ plane, and the
magnetic field -- the positive $z$-axis. The incident light comes from the
$z<0$ subspace, forming an angle $\phi_i$ with the $z$ axis. The 
$\vec q_i,~\vec B$ pair of vectors define the $xz$ plane, i. e. the 
projection $\vec q_{i\Vert}$ is oriented along the positive $x$ axis. The
emitted light goes back to the $z<0$ subspace. It is characterized by 
angles $\phi_f$ with the $z$-axis, and $\theta_f$ in the $xy$ plane. We
will take, for the incident light:

\begin{eqnarray}
q_{i\Vert}&=& q_i \sin \phi_i, \nonumber\\
\varepsilon_{iy}&=&1,
\end{eqnarray}

\noindent
whereas for the scattered light: $q_{f\Vert}=q_f \sin \phi_f$.
We will distinguish two situations: (i) The ``parallel'' light polarization,
in which:

\begin{eqnarray}
\varepsilon_{fx}&=&-\sin \theta_f,\nonumber\\
\varepsilon_{fy}&=&\cos \theta_f,
\end{eqnarray}

\noindent
and (ii) The ``perpendicular'' light polarization, where:

\begin{eqnarray}
\varepsilon_{fx}&=&\cos \phi_f \cos \theta_f,\nonumber\\
\varepsilon_{fy}&=&\cos \phi_f \sin \theta_f.
\end{eqnarray}

Below, we give an explicit expression for the matrix element of the 
$H_{e-r}^-$ hamiltonian:

\begin{equation}
_{N+1}\langle int|H_{e-r}^-|i\rangle_N=\frac{e}{m_0}\sqrt{\frac{2\pi\hbar}
 {V\omega_i\eta_i^2}}\sum_{\sigma,\tau} band_{\sigma,\tau}^{(i)}\times
 orbital^{(i)}_{\sigma,\tau} \times V_{\sigma,\tau}^{(N+1)*}.
\label{Rayleigh}
\end{equation}

The band and orbital factors are evaluated in Appendix \ref{Appendix A}. 
The energy denominator in the scattering amplitude will be written in
the form:

\begin{equation}
\hbar\omega_i-(E_{int}-E_i)=\hbar\omega_i-E_{gap}-\hbar
 \Omega_{int}^{(N+1)},
\label{denominator}
\end{equation}

\noindent
where $\hbar\Omega_{int}$ is the eigenvalue coming from the pp-TDA
equations, and $E_{gap}=1560$ meV is a nominal band gap.

The matrix element of the $H^+_{e-r}$ operator, has
a bit more cumbersome expression:

\begin{eqnarray}
_N\langle f|H^+_{e-r}|int\rangle_{N+1}=&-&\frac{e}{m_0}
 \sqrt{\frac{2\pi\hbar}{V\omega_f\eta_f^2}}~\sum_{\sigma,\tau}
 \sum_{\sigma',\lambda} \delta(\sigma,\sigma')~band_{\lambda\tau}^{(f)}
 \times orbital^{(f)}_{\lambda\tau}\times V_{\sigma\tau}^{(N+1)}
 X_{\sigma'\lambda}^{e*}\nonumber\\
&-& \frac{e}{m_0}
 \sqrt{\frac{2\pi\hbar}{V\omega_f\eta_f^2}}~
 \sum_{\sigma,\tau}\sum_{\tau',\lambda} \delta(\tau,\tau')~
 band_{\sigma\lambda}^{(f)}\times orbital^{(f)}_{\sigma\lambda} \times
 V_{\sigma\tau}^{(N+1)} X^{h*}_{\tau'\lambda}.
\end{eqnarray}

\noindent
The interpretation is, however, straightforward. Let us take the first
term. The first sum runs over e-h states $(\sigma,\tau)$, both above the
Fermi levels, entering the pp-TDA function $|int\rangle_{N+1}$. 
$V_{\sigma\tau}^{(N+1)}$ are the corresponding coefficients. The second 
sum represents the electronic excitation part of the RPA function 
$|f\rangle_N$. $\lambda$ is an electronic state below $\mu_F^e$, and 
$\sigma'$ an state above $\mu_F^e$. The transition from 
$|int\rangle_{N+1}$ to $|f\rangle_N$ is caused by a pair of annihilation
operators $he$. It is evident that the subindexes should be 
$h_{\tau}e_{\lambda}$, and thus $\sigma'=\sigma$.

Amplitudes for backscattering processes, in which $\phi_f=\phi_i$, and
$\theta_f=\pi$, will be computed.

\section{Results}
\subsection{Properties of the HF ground state}

We give in this sub-section a few qualitative results that follow from the 
HF calculations.

We show in Fig. \ref{fig2} a subset of the HF single-particle 
levels at $B=1$ T. Fermi energies are represented as dotted lines. Apart
from an overall downward shift, we observe only a slight deformation of free
oscillator shells due to Coulomb interactions. The Zeeman splitting is not 
resolved in the figure scale, thus spin-up and -down levels are 
simultaneously occupied. As a result, total electron and hole spins remain 
equal to zero when the magnetic field is varied between zero and 2 T. In
fact, very low spin polarizations persist up to higher magnetic fields,
of the order of 20 T. \cite{Boris y yo} The total ground-state angular 
momentum is also zero in this magnetic field range, and persists up to very
high $B$, as a prelude to the formation of e-h pairs in zero relative
angular momentum states, which maximize Coulomb attraction.

Let us stress also that Fig. \ref{fig2} qualitatively predicts that 
single-particle excitations (SPE) with $\Delta L_z=\pm 2$ 
(``quadrupole''-like, represented by arrows in the figure) are lower in
energy than ``monopole'' ($\Delta L_z=0$) or ``dipole'' 
($\Delta L_z=\pm 1$) excitations at $B=1$ T. This fact is corroborated by 
the RPA calculations, see below. The energetic cost of adding an e-h pair
with $l_e+l_h=0$ is, according to Fig. \ref{fig2}, around 15 meV (plus
$E_{gap}$). A value confirmed by the TDA results.

Electron and hole densities at $B=0$ and 2 T are drawn in Fig. \ref{fig3}. 
The small differences between both densities are due to the differences
between electron and hole in-plane masses. The maximum value, around 
$7\times 10^{11}$ pairs/cm$^2$, is typical of excitonic systems. 
Notice also that density oscillations, related to 
shell-filling effects, are smoothed as $B$ is increased. This fact is due
to the increasing occupations of states in the first Landau level, which
wave functions have no radial nodes.

\subsection{Multipole excitations and renormalization of the absorption 
edge}

CDE of various multipolarities in the $N$-pair system are obtained from 
the RPA computations. We show in Fig. \ref{fig4} the monopole sector,
which is the most relevant for Raman CDE processes. States with more than
5 \% contribution to the energy-weighted sum rule (\ref{S1}) are 
represented in the figure as triangles. They will be called ``collective''
excitations. They form three well defined bands accounting for, 
approximately, 7, 35 and 5\% of the sum rule. The rest of the monopole 
strength is divided among 200 states with excitation energy lower than 
30 meV. Fig. \ref{fig4} shows also the lowest monopole SPE (triangles plus
dotted line).
A complex pattern of probability transfer between colliding levels, as the
magnetic field is varied, is reflected in Fig. \ref{fig4} in the form of 
abrupt variations of the number of collective levels. The situation is 
similar to the behavior of the dipole strength in the biexciton 
\cite{Ricardo y yo}.

Dipole and quadrupole collective levels and the corresponding SPE in these
sectors are shown in Fig. \ref{fig5}. One sees that dipole excitations are,
as a rule, lower than monopole and quadrupole collective CDE, but the
quadrupole SPE are lower at $B=1$ T, as mentioned above with regard to 
Fig. \ref{fig2}.

We show in Fig. \ref{fig6} a few results following from the pp-TDA calculations.
In the upper figure, the lowest $\hbar\Omega_{int}$ for the intermediate 
state with $N+1$ pairs, in which the added pair has $l_e+l_h=0$, is drawn. 
This magnitude can be taken as the renormalization of the absorption edge due
to the background of $N$ electron-hole pairs. Let us stress that there are two
main effects contributing to this magnitude. The first is the blue shift
induced by Fermi statistics, i. e. the added pair should occupy higher 
HF single-particle states. The second is the red shift caused by Coulomb
(attractive)
interactions. As can be seen in this figure, for $N=42$ the net result is a blue
shift of 12 - 16 meV. The apparent kinks are signals of ground-state rearrangements
as the magnetic field is varied.

The lowest part of Fig. \ref{fig6} shows the dependence on $N$ of the edge
renormalization at $B=1$ T. It grows from 3 meV for 12 pairs up to 14 meV 
for the 42-pair system. That is, at a rate of 0.3 meV per pair in
the dot. This magnitude can be used as a complementary way of determining
the mean number of pairs in the dot.

\subsection{Raman spectra in CDE channels}

Let us consider Raman processes in which the final states are CDE. The first
important question we would like to address is the role played by 
collective and SPE in resonant Raman amplitudes. 

We show in Fig. \ref{fig7} the Raman differential cross section, computed
from

\begin{equation}
\frac{{\rm d}^2\sigma}{{\rm d}\Lambda_f{\rm d}\omega_f}=
\frac{V^2 \omega_f^3 \eta_f \eta_i^3}{4 \pi^2 c^4 \omega_i \hbar N_i} \sum_f
|A_{fi}|^2~ \delta (E_i+\hbar\omega_i-E_f-\hbar\omega_f),
\end{equation}

\noindent
in which ${\rm d}\Lambda_f$ is the solid angle element in the direction of the
dispersed light. We will use a smearing of the delta function as follows:

\begin{equation}
\delta (x)=\frac{\Gamma_f/\pi}{x^2+\Gamma_f^2},
\end{equation}

\noindent
with a phenomenological $\Gamma_f=0.5$ meV.

The spectra in Fig. \ref{fig7} are computed under conditions of normal incidence
($\phi_i=\phi_f=0$, only monopole final states are excited) and 
parallel light polarization. The latter is supposed to disentangle collective
CDE modes from SDE in electronic qdots under non-resonant scattering
\cite{Ando}. The monopole strengths are also included in the figure for comparison.

The upper figure shows results at $B=1$ T. There are always Raman peaks
associated to the more collective CDE states,
although their magnitude rapidly vary with the incident laser frequency. The
overall behavior, which is apparent in this figure, is that low-energy
weakly collective or SPE are favored at ``extreme resonance'', i. e. when
$\hbar\omega_i$ is near 1574 meV in this situation, whereas 30 meV above
the effective band gap high-energy collective or weakly collective CDE
give the strongest peaks.

Notice that the maximum peak intensities under extreme resonance are reached 
for laser frequencies a few meV above the renormalized band gap.
Thus, as the laser frequency moves above 1574 meV, the amplitudes 
corresponding to weakly collective or SPE initially increase, but further 
experience a sudden drop. 

In the lower part of Fig. \ref{fig7}, the spectra at $B=2$ T are drawn. 
We notice variations in the peak distributions as compared to the $B=1$ T
results. Notice also that, in both spectra, the strongest CDE state is not 
seen as a distinct peak for $\hbar\omega_i=1600$ meV.

The qualitative conclusions to be extracted from Fig. \ref{fig7} are thus
the following: (a) Comparable Raman intensities for strongly collective and 
for weakly collective states (even for SPE at intermediate excitation energies),
(b) A complex pattern of variations of the Raman intensities as the frequency of 
the incident laser or the magnetic field is varied. Distinct peaks are seen 
only in certain intervals of these magnitudes, and (c) A richer structure of 
the Raman spectra as compared with the charged quantum dots.

The second important point to discuss, from the qualitative point of view,
is the excitation of high multipolarity modes at non-zero momentum transfer.
We notice that the light wave vector is around $8\times 10^3$ nm$^{-1}$ in
the present situation. The maximum momentum transfer is thus $1.6\times 10^4$
nm$^{-1}$ in backscattering geometry. We show in Fig. \ref{fig8} the quadrupole
spectra at $B=1$ T in the parallel light polarization configuration and 
momentum transfer equal to $8\times 10^3$ nm$^{-1}$ ($\phi_i=\phi_f=\pi/6$).
Of course, in an experimental curve all the multipolarities come together. We
separate the quadrupole spectra to simplify the analysis.

First, we notice that quadrupole Raman intensities are 1/10 of monopole ones.
On qualitative grounds, one expects quadrupole intensities of order
$N_{states}^2 (q_{i\Vert} D)^4$, where $D\approx 90$ nm is the system
diameter, and $N_{states}$ is the number of intermediate states participating
in the process. $(q_{i\Vert} D)^4$ provides a factor $10^{-2}$, but the 
number of states contributing to quadrupole processes is roughly three times
the states contributing to monopole processes (intermediate states with
excess angular momentum 0, +1, and +2 in the $\Delta l=+2$ case, for
example). Thus $N_{states}^2 (q_{i\Vert} D)^4\approx 10^{-1}$. Second, we
observe an asymmetry between the $\Delta l=-2$ and $\Delta l=2$ spectra. 
Most of the $\Delta l=-2$ peaks correspond to SPE or weakly collective states.
The $\Delta l=2$ peaks, two or three times more intense, are concentrated around 
collective states, which strengths are more uniformly distributed. The
most collective CDE state with $\Delta l=2$ shows up as a distinct peak only
in a thin range of frequencies. Other multipoles show similar behavior.

Thus, the conclusions coming from Fig. \ref{fig8} are the following: 
(a) The intensity of CDE Raman peaks with multipolarity $l$ are 
proportional to $(|l|+1)^2 (q_{i\Vert} D)^{2 |l|}$, and (b) Negative-$l$
peaks correspond mainly to very weakly collective or SPE states. The peak
associated to the most
collective CDE state is well defined practically at any $\hbar\omega_i$.
On the other hand, positive-$l$ peaks are stronger and show a dominance of
collective states.

A third interesting question to be addressed is related to the modes excited
when the dispersed light polarization is orthogonal to the polarization of the
incident light. The results for monopole states at $B=1$ T are presented in
Fig. \ref{fig9}. Under extreme resonance, we observe peaks associated to
SPE modes with excitation energies lower than 12 meV. In particular, the 
lowest SPE at 4 meV is clearly distinguished. Raman signals due to CDE 
states are strongly suppressed in these conditions. 30 meV above extreme 
resonance, the dominant peaks are located at higher excitation energies. They
correspond to SPE or weakly collective states.

\section{Conclusions}

We have computed the Raman amplitudes for the light scattered of a qdot 
which contains 42 e-h pairs. In an attempt to identify the states giving
rise to the strongest peaks, we compared the Raman intensities with the
multipole strengths. The result is that both collective and SPE states
play important roles in Raman spectra. Their relative weight in the 
spectra is seen to strongly depend on the external magnetic field, the
polarization of the scattered light and the frequency of the incident light.
Taken in a wider context, this conclusion suggests caution when making
an assignment to an experimental Raman peak, and urges for theoretical
calculations in parallel to the experiments.

The explicit construction of the wave functions for the intermediate
states, always in the framework of mean-field time-dependent 
approximations, allows us to consider extreme as well as
non extreme resonance conditions. In the same way, the formalism allows for
any wave momentum transfer or any kind of light polarization.

We can not presently consider SDE final states. The reason
has been explained briefly in the text: the SDEs are ``2p-2h'' states,
which can not be modeled by the RPA approximation adopted in 
this work. It shall be said, however, that none of the papers 
\onlinecite{Steinebach} or \onlinecite{Emperador} accounted
for valence-band mixing effects in the intermediate hole state
in Raman SDE channels.

However, on qualitative grounds, it can be argued that SDE final states
shall give strong Raman peaks, may be even stronger than CDE states. 
The argument goes as follows. It may be seen that the factor determining 
the Raman amplitude is in fact the orbital factor, i. e. the overlapping
between the electron and hole wave functions. In the symmetric, $N_e=N_h$,
system we are studying, the overlapping is high in the intermediate states
(both e and h above the Fermi levels), but low for CDE final states 
because one of the annihilated particles is above its Fermi level, and the
other is below. For SDE states, however, both the annihilated e and h are
below their Fermi levels, and the overlapping may be high. Thus, Raman
SDE amplitudes could be even stronger than CDE amplitudes.

In the electronic qdots, Raman scattering in SDE channels goes through hole 
band mixing. Apart from the low overlapping in final states, one would expect 
the amplitude to be proportional to the light hole component of the
hole wave function. Due to the fact that the Coulomb interaction is diagonal 
in the band indexes, a strong electronic background could depress valence
band mixing, thus making SDE amplitudes even weaker. Research along this
direction is in progress.

\acknowledgements
Part of this work was carried out at the Abdus Salam ICTP. A. D. and A. G.
acknowledge the ICTP Associate and Federation Schemes Office for support.
 
\appendix
\section{Evaluation of band and orbital factors}
\label{Appendix A}

We give in this Appendix the expressions for the band and orbital factors
entering Eqs. (\ref{Her-}) and (\ref{Her+}). The ratio of the band factor
$\vec\varepsilon_i\cdot\vec p_{\alpha\gamma}$ to the magnitude $iP$, where 
$P$ is the interband GaAs constant, is given in Tab. \ref{tab1}. 
$S_z=\pm 1/2$ is the spin projection over the $z$ axis.

Conventionally, we assign $S_z^h=-1/2$ to the $m_j=3/2$ electron
state in the valence band. The $\varepsilon_{\pm}$ components are defined as

\begin{equation}
\varepsilon_{\pm}=\mp \frac{\varepsilon_x\mp i \varepsilon_y}{\sqrt{2}}.
\end{equation}

The band factor entering Eq. (\ref{Her+}), i. e. 
$\vec\varepsilon_f^{~*}\cdot\vec p^{~*}_{\alpha\gamma}$, 
can be also obtained  
from Table \ref{tab1} if we replace $\varepsilon_+$ by $\varepsilon_-$
and viceversa.

On the other hand, the orbital factor in Eq. (\ref{Her-}) is computed 
from the HF one-particle functions, Eq. (\ref{expansion}). Substituting
(\ref{expansion}) into the expression for the band factor, Eq. (\ref{Her-}),
and making use of the expansion:

\begin{equation}
e^{i \vec q_i\cdot\vec r}\approx 1+i \vec q_{i\Vert} \cdot\vec r_{\Vert}-
 \frac{1}{2} (\vec q_{i\Vert} \cdot\vec r_{\Vert})^2, 
\end{equation}

\noindent
where $\vec r_{\Vert}$ means the projection of $\vec r$ onto the $xy$
plane, we get

\begin{eqnarray}
\int &e^{i \vec q_i\cdot\vec r}&\phi^*_{\alpha e}(\vec r)
 \phi^*_{\gamma h}(\vec r) {\rm d}^3 r\approx
 \sum_{s,t} C^{e*}_{\alpha s} C^{h*}_{\gamma t} \left\{
 \langle k_s,l_s|1-\frac{q_{i\Vert}^2}{4} d_0|k_t,-l_t\rangle \right.+\nonumber\\
 &i& \frac{q_{i\Vert}}{2}\langle k_s,l_s|e^{-i\theta_i} d_1
 +e^{i\theta_i} d_{-1}|k_t,-l_t\rangle \nonumber\\
 &-& \frac{(q_{i\Vert})^2}{8} \left.\langle k_s,l_s|e^{-2 i\theta_i} 
 d_2+ e^{2 i\theta_i} d_{-2}|k_t,-l_t \rangle \right\}.
\label{orbital}
\end{eqnarray}

In this last equation, $d_l$ are the one-particle multipole
operators, which explicit expression is given in Appendix \ref{Appendix B}.
The  $k_s$ and $l_s$ are, respectively, the radial and orbital
quantum numbers of the 2D oscillator state $\chi_s$. The angle $\theta_i$
is by definition equal to zero, i. e. the $z$ axis is oriented along
$\vec q_{i\Vert}$.

The orbital factor entering Eq. (\ref{Her+}) can be obtained formally
from (\ref{orbital}) upon substituting $i$ by $f$ and taking the complex
conjugate of the whole expression.

\section{Multipole matrix elements and sum rules}
\label{Appendix B}

The evaluation of one-particle elements, $d^l_{\alpha,\gamma}$, requires
the expansion (\ref{expansion}) for HF functions:

\begin{equation}
d^l_{\alpha,\gamma}=\sum_{s,t} C^*_{\alpha s} C_{\gamma t}
 d^l_{st},
\end{equation}

\noindent
where the elements $d^l_{st}$, taken over oscillator functions, are given,
when $l=0$, by:

\begin{eqnarray}
\langle k_s,l_s|\frac{d_0}{l_0^2}|k_t,l_t\rangle&=&\delta(l_s,l_t)\left\{ 
 (2 k_t+|l_t|+1)\delta(k_s,k_t)-\sqrt{(k_t+1)(k_t+|l_t|+1)}~
 \delta(k_s,k_t+1)\right.\nonumber\\
 &-&\left. \sqrt{k_t (k_t+|l_t|)}~\delta(k_s,k_t-1)\right\}.
\end{eqnarray}

Whereas, for $l>0$:

\begin{eqnarray}
\langle k_s,l_s|\frac{d_l}{l_0^l}|k_t,l_t\rangle&=&\delta(l_s,l_t+l)
 \sum_{r=0}^{{\rm Min}(l,k_t)} (-1)^r \frac{l!}{(l-r)! r!}
 \sqrt{\frac{k_t!}{(k_t+|l_t|)!}\frac{(k_t-r+|l_t+l|)!}{(k_t-r)!}}
 \nonumber\\
 &&\times \delta(k_s,k_t-r);~~~l_t\ge 0,\nonumber\\
&=&\delta(l_s,l_t+l)
 \sum_{r=0}^{{\rm Min}(l,k_s)} (-1)^r \frac{l!}{(l-r)! r!}
 \sqrt{\frac{k_s!}{(k_s+|l_t+l|)!}\frac{(k_s-r+|l_t|)!}{(k_s-r)!}}
 \nonumber\\
 &&\times \delta(k_t,k_s-r);~~~l_t\le -l,\nonumber\\
&=&\delta(l_s,l_t+l) (-1)^{k_s-k_t}
 \sum_{r=0}^{{\rm Min}(l-|l_t|,k_t)}\frac{(l-|l_t|)!}{(l-|l_t|-r)! r!}
 \frac{|l_t|!}{(|l_t|-k_s+k_t-r)!}\nonumber\\
 &&\times\sqrt{\frac{k_t!}{(k_t+|l_t|)!}\frac{k_s!}{(k_s+|l_t+l|)!}
 \frac{(k_t-r+l)!^2}{(k_t-r)!^2}}\Theta(k_s-k_t+r);\nonumber\\
&&-l<l_t<0.
\end{eqnarray}

\noindent
Where $\Theta(x)=1$ for $0\le x\le {\rm Min}(|l_t|,k_s)$ and zero
otherwise. Finally, for $l<0$, we get:

\begin{equation}
\langle k_s,l_s|d_l|k_t,l_t\rangle=
 \langle k_t,l_t|d_{|l|}|k_s,l_s\rangle^*.
\end{equation}

On the other hand, the elements 
$\langle \lambda|(r^2)^{\xi}|\lambda \rangle$, entering the r. h. s. of
the sum-rule equations (\ref{S1}-\ref{S10}) are evaluated as:

\begin{equation}
\langle \lambda|(r^2)^{\xi}|\lambda \rangle=\sum_{s,t}
 C^*_{\lambda s} C_{\lambda t} \langle s|(r^2)^{\xi}|t \rangle,
\end{equation}

\noindent
where:

\begin{eqnarray}
\langle k_s,l_s|\left(\frac{r^2}{l_0^2}\right)^{\xi}|k_t,l_t \rangle&=&
 \delta(l_s,l_t)
 \sqrt{\frac{k_s! k_t!}{(k_s+|l_s|)!(k_t+|l_s|)!}}
 \sum_{m=0}^{{\rm Min}(\xi,k_s)}\sum_{n=0}^{{\rm Min}(\xi,k_t)}
 (-1)^{m+n}\nonumber\\
&&\times \delta(k_s-m,k_t-n)\frac{\xi!^2}{(\xi-m)! m! (\xi-n)! n!}
 \frac{(k_t-n+|l_s|+\xi)!}{(k_t-n)!}.
\end{eqnarray}

\begin{table}
\begin{tabular}{|l||l|l|}
$S_{z\alpha }^e \backslash ~S_{z\gamma }^h $& -1/2 & 1/2 \\ 
\hline\hline
1/2 & $\varepsilon _{+i}$ & 0 \\ 
\hline
-1/2 & 0 & $\varepsilon _{-i}$
\end{tabular}
\caption{The quotient $\vec\varepsilon_i\cdot\vec p_{\alpha\gamma}/(iP)$.}
\label{tab1}
\end{table}

\begin{figure}
\caption{(a) Inelastic light scattering leading to final states 
 which are CDE of the ground state. (b) An example of Raman scattering 
 in SDE channels.}
\label{fig1}
\end{figure}

\begin{figure}
\caption{A set of electron and hole HF levels at $B=1$ T. The Fermi 
energies are indicated as dotted lines. The less energetic transitions 
with $\Delta L_z=\pm 2$ are represented by arrows.}
\label{fig2}
\end{figure}

\begin{figure}
\caption{Electron and hole densities in the HF ground state.}
\label{fig3}
\end{figure}

\begin{figure}
\caption{Monopole collective CDEs of the $N$-pair system and the lowest 
SPE.}
\label{fig4}
\end{figure}

\begin{figure}
\caption{Dipole and quadrupole collective CDEs and the lowest SPEs.}
\label{fig5}
\end{figure}

\begin{figure}
\caption{Absorption edge renormalization: (a) As a function of $B$
for $N=42$, and (b) As a function of $N$ for $B=1$ T.}
\label{fig6}
\end{figure}

\begin{figure}
\caption{Raman spectra under conditions of normal incidence and 
parallel polarization. Only final CDE states are considered.}
\label{fig7}
\end{figure}

\begin{figure}
\caption{Raman spectra of quadrupole CDE states. $\phi_i=\phi_f=\pi/6$ and
 parallel light polarization geometry.}
\label{fig8}
\end{figure}

\begin{figure}
\caption{The same as in Fig. 7, but the polarization of the scattered light is 
orthogonal to the polarization of the incident light.}
\label{fig9}
\end{figure}


\begin{references}
\bibitem[a]{alain} Electronic address:
 gran@ceaden.edu.cu
\bibitem[b]{augusto} Electronic address: 
 agonzale@cidet.icmf.inf.cu
\bibitem[c]{eduardo} Electronic address:
 eariel@raman.ff.oc.uh.cu
\bibitem{Lockwood} D. J. Lockwood, P. Hawrylak, P. D. Wang et. al., Phys.
 Rev. Lett. {\bf 77}, 354 (1996).
\bibitem{Schuller} C. Schuller, K. Keller, G. Biese et. al., Phys. Rev.
 Lett. {\bf 80}, 2673 (1998).
\bibitem{Steinebach} C. Steinebach, C. Schuller, and D. Heitmann, Phys.
 Rev. {\bf B 59}, 10240 (1999).
\bibitem{Emperador} M. Barranco, L. Colleti, E. Lipparini et. al., 
 Phys. Rev. {\bf B 61}, 8289 (2000).
\bibitem{Alain} A. Delgado and A. Gonzalez, to be submitted.
\bibitem{Wolfe} J. C. Kim and J. P. Wolfe, Phys. Rev. B {\bf 57}, 9861 
 (1998).
\bibitem{WHFJ96} A. Wojs, P. Hawrylak, S. Fafard, and 
 L. Jacak, Phys. Rev. {\bf B 54}, 5604 (1996).
\bibitem{Ring-Schuck} P. Ring and P. Schuck, The nuclear many-body
 problem, Springer-Verlag, New-York (1980). 
\bibitem{Trallero} E. Men\'endez-Proupin, C. Trallero-Giner and S. E. Ulloa,
 Phys. Rev. {\bf B 60}, 16747 (1999). 
\bibitem{Eduardo} E. Men\'endez-Proupin, Inelastic light scattering in semiconductor
 quantum dots, Ph. D. Thesis, Havana University (2000).
\bibitem{Davies} H. D. M. Davies, J. C. Harris, J. F. Ryan, and A. J. 
 Turberfield, Phys. Rev. Lett. {\bf 78}, 4095 (1997).
\bibitem{gdr} A. Delgado, L. Lavin, R. Capote and A. Gonzalez, Physica E
 {\bf 8}, 342 (2000).
\bibitem{Nickel} H. A. Nickel, T. Yeo, C. J. Meining et. al., cond-mat/0106277.
\bibitem{Boris y yo} B. A. Rodriguez and A. Gonzalez, Phys. Rev. B {\bf 63} 
 205324 (2001).
\bibitem{tesis Alain} A. Delgado, Giant dipole resonances in quantum
 dots, M. Sc. Thesis, Inst. for Nuclear Sci. and Tech., Havana (2000).
\bibitem{Boris} We use a parametrization by B. Rodriguez of the 
 experimental $g$ values for quantum wells given in S. P. Najda, 
 S. Takeyama, N. Miura et. al., Phys. Rev. B {\bf 40}, 6189 (1989);
 M.J. Snelling, G. P. Flinn, A.S. Plaut, R. T. Harley, A. C.
 Tropper, R. Eccleston, and C. C. Phillips, Phys. Rev. B {\bf 44}, 11345
 (1991); M. J. Snelling, E. Blackwood, C. J. McDonagh, R. T. Harley,
 and C. T. B. Foxon, Phys. Rev. B {\bf 45}, 3922 (1992);
 N. J. Traynor, R. J. Warburton, M. J. Snelling, and R. T.
 Harley, Phys. Rev. B {\bf 55}, 15701 (1997);
 M. Seck, M. Potemski, and P. Wyder, Phys. Rev. B {\bf 56},
 7422 (1997).
\bibitem{MacD} A. H. MacDonald and G. C. Aers, Phys. Rev. {\bf B 29},
 5976 (1984).
\bibitem{Ricardo y yo} R. Perez and A. Gonzalez, J. Phys.: Condens. Matter
 {\bf 13}, L539 (2001).
\bibitem{Ando} S. Katayama and T. Ando, J. Phys. Soc. Jpn. {\bf 54}, 1615
 (1985).
\end{references}
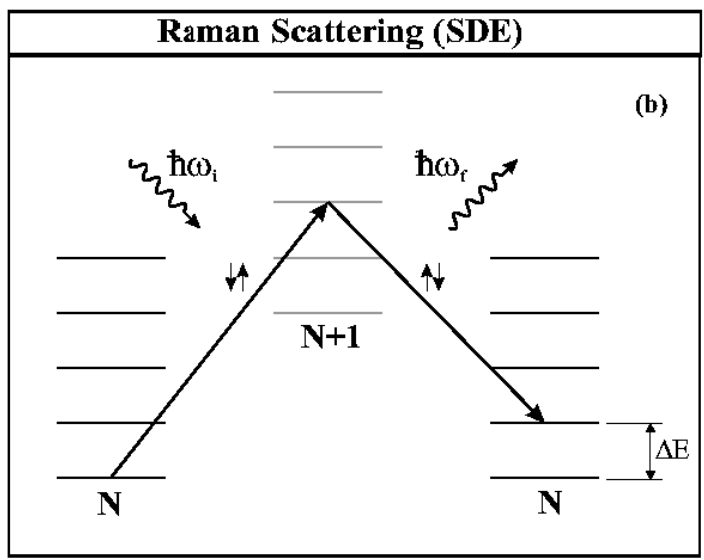
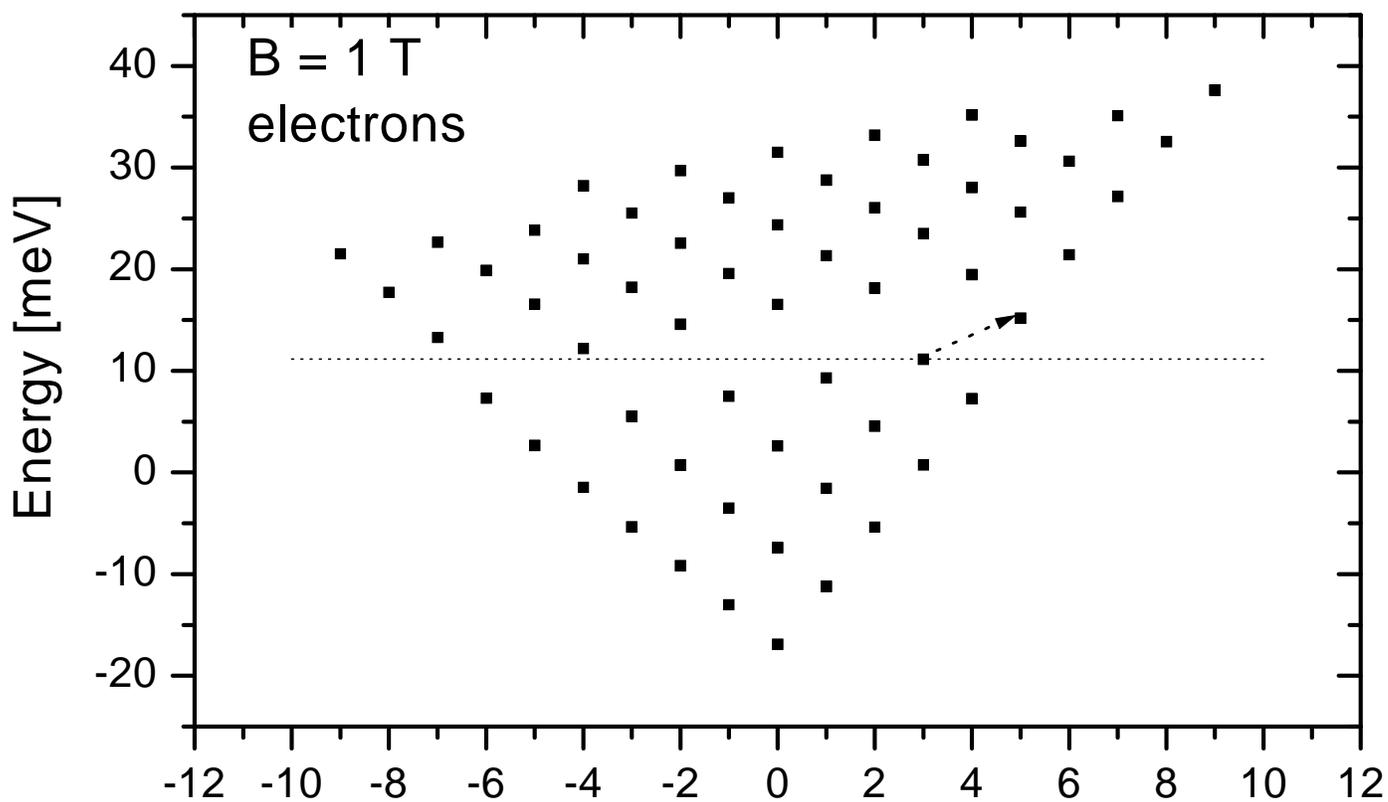
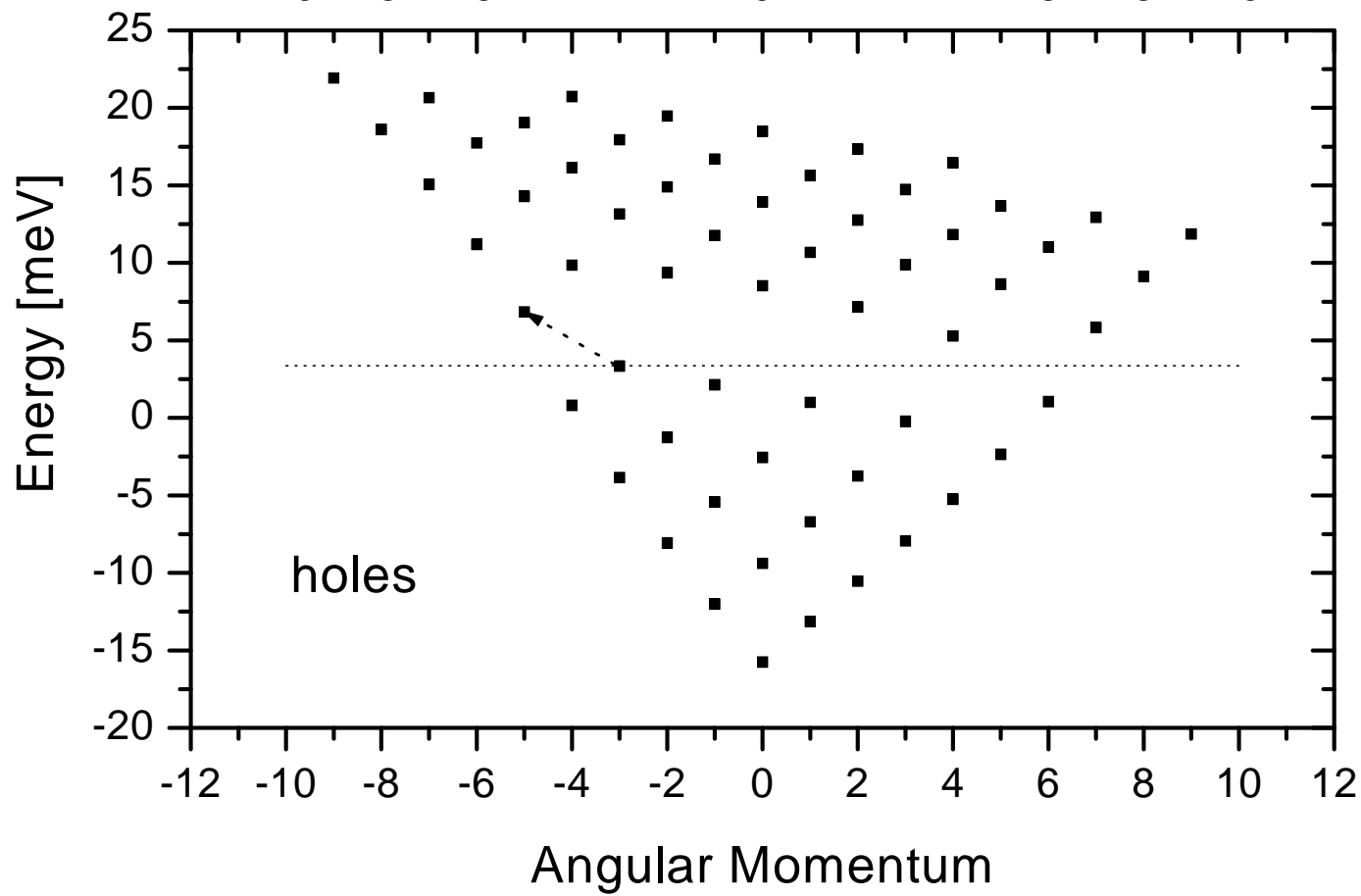
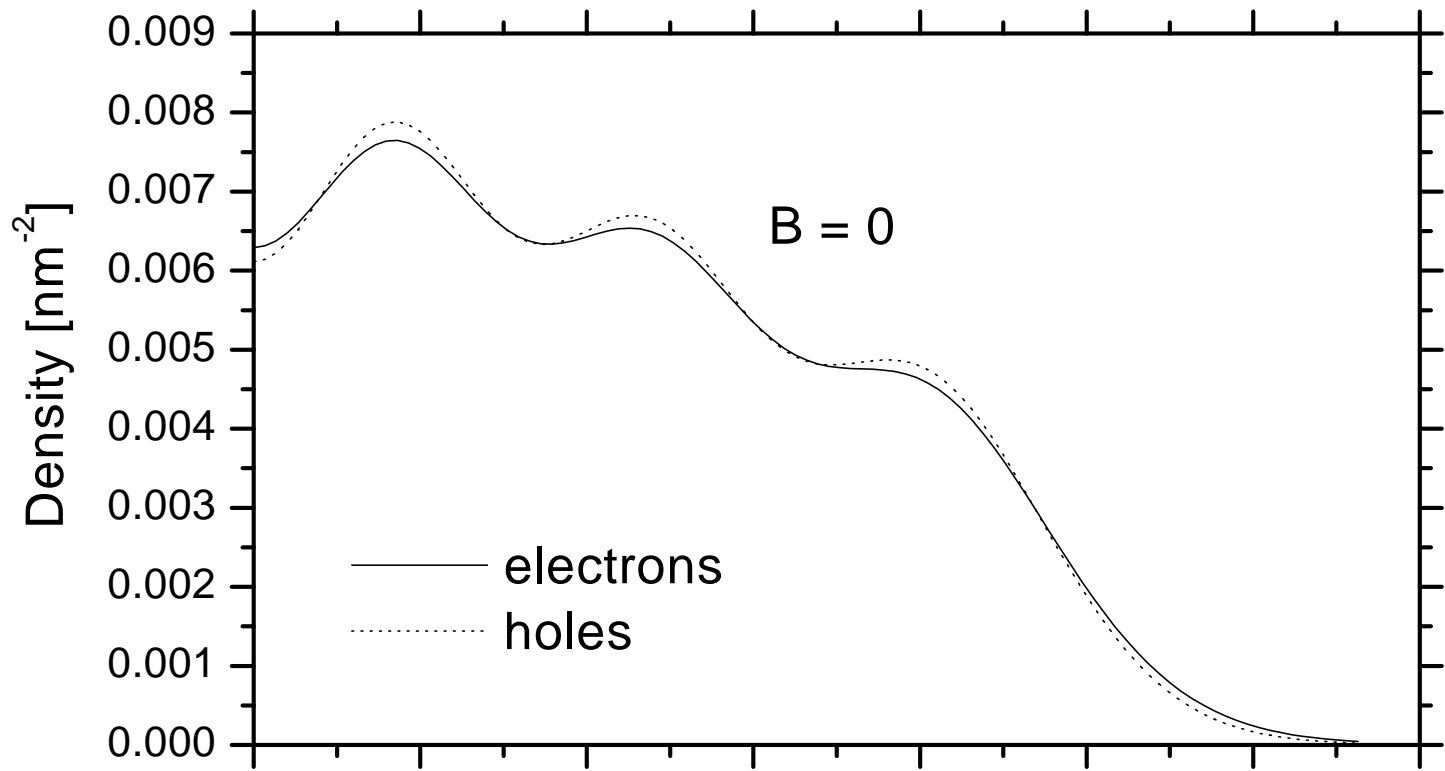
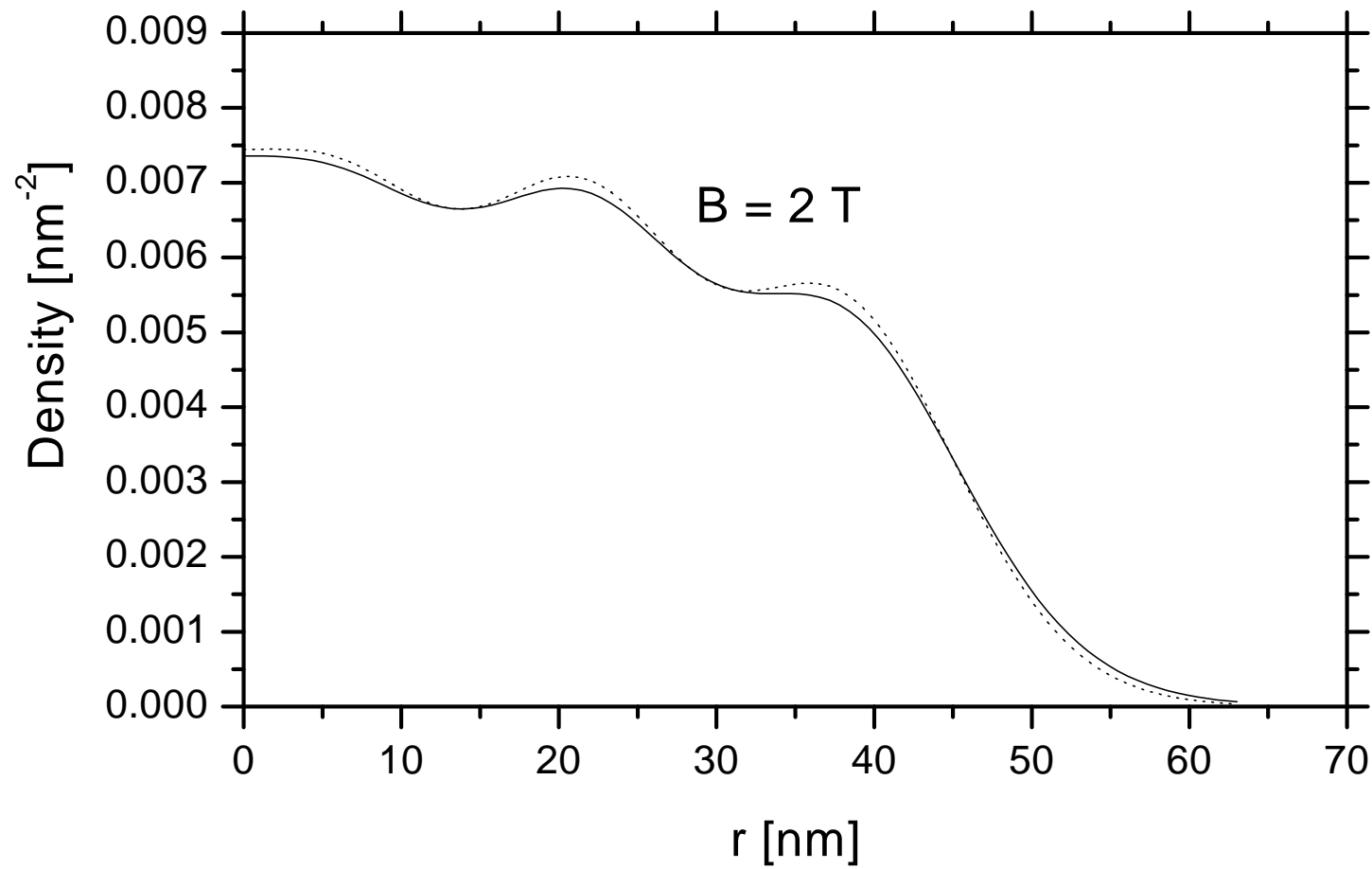
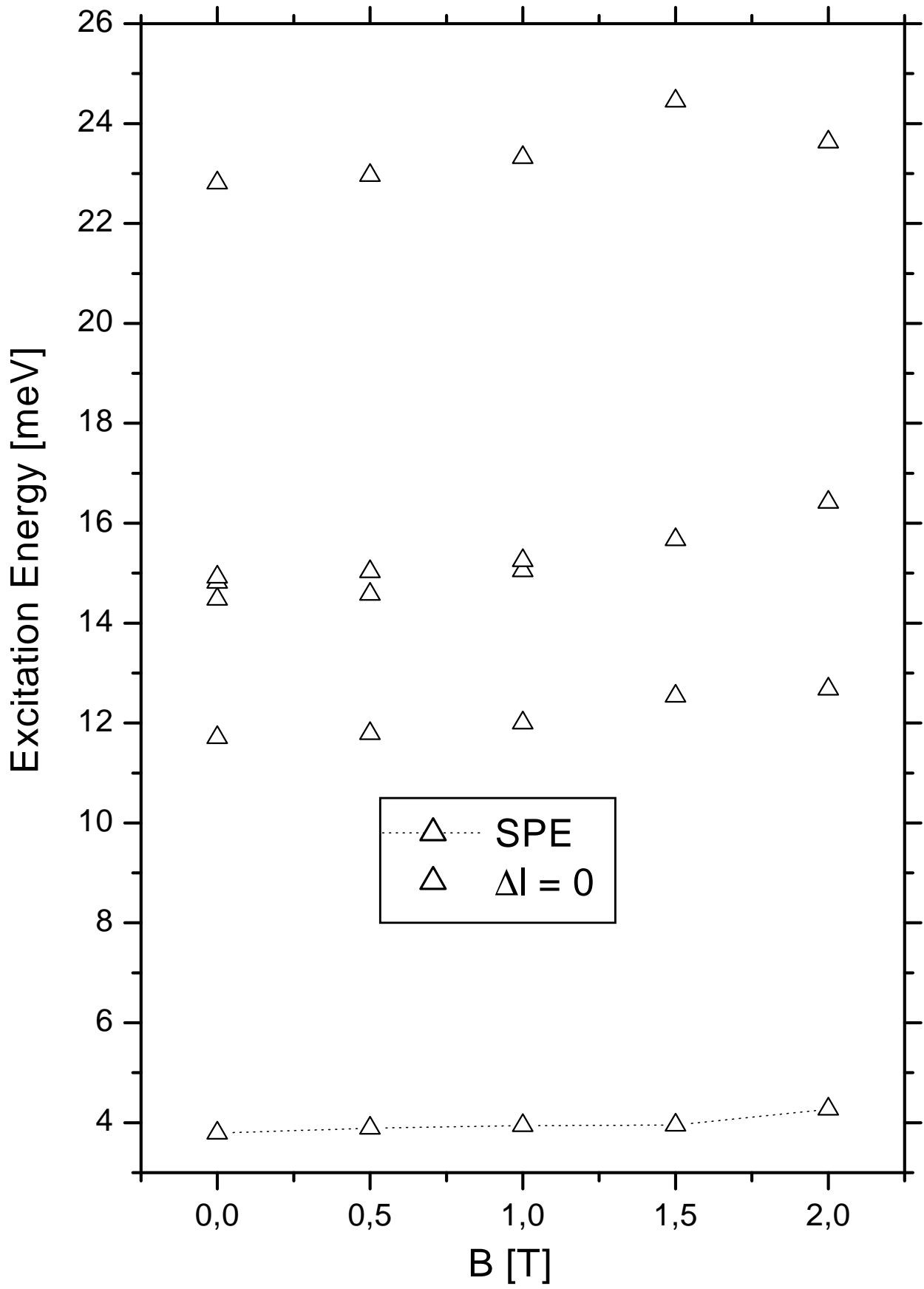
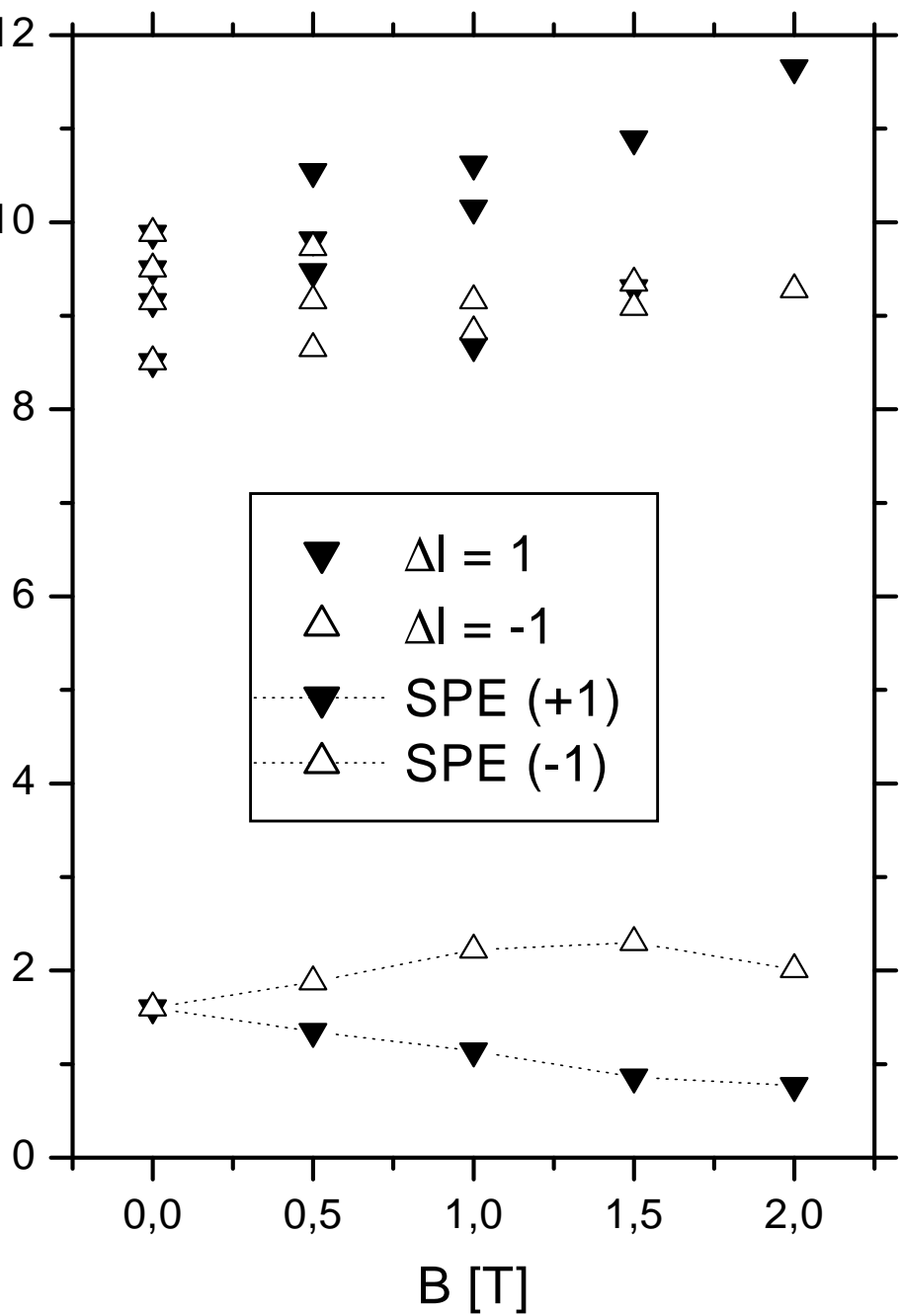
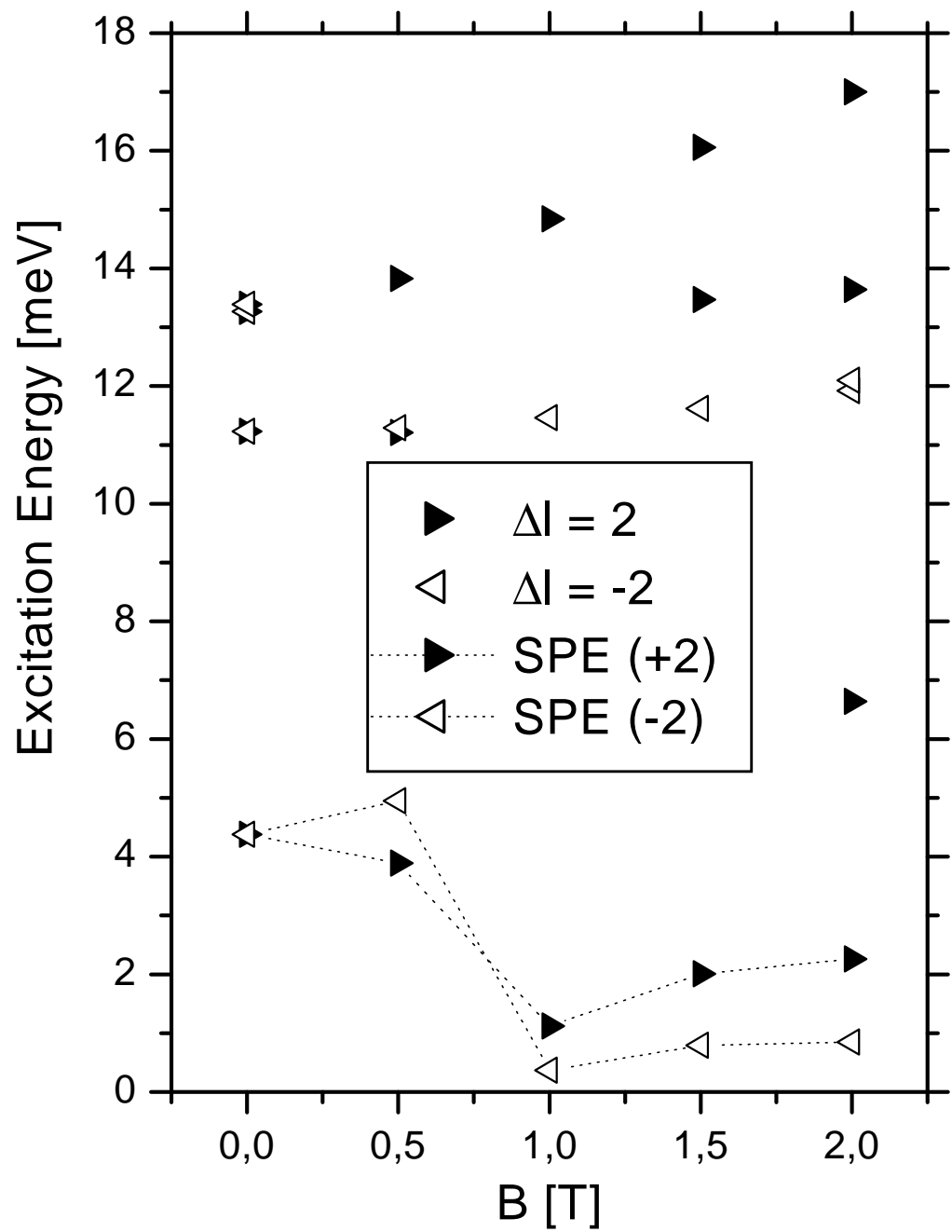
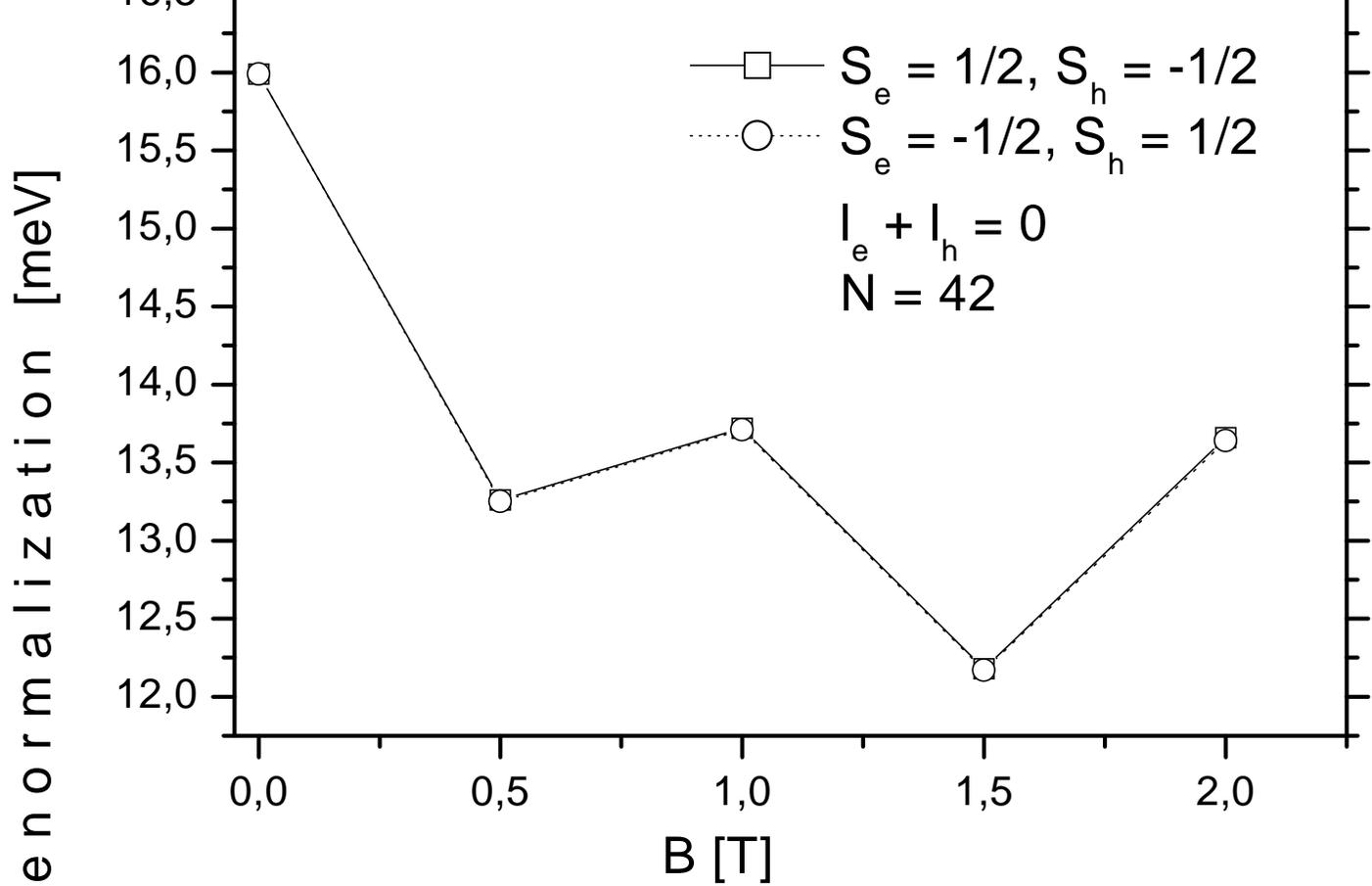
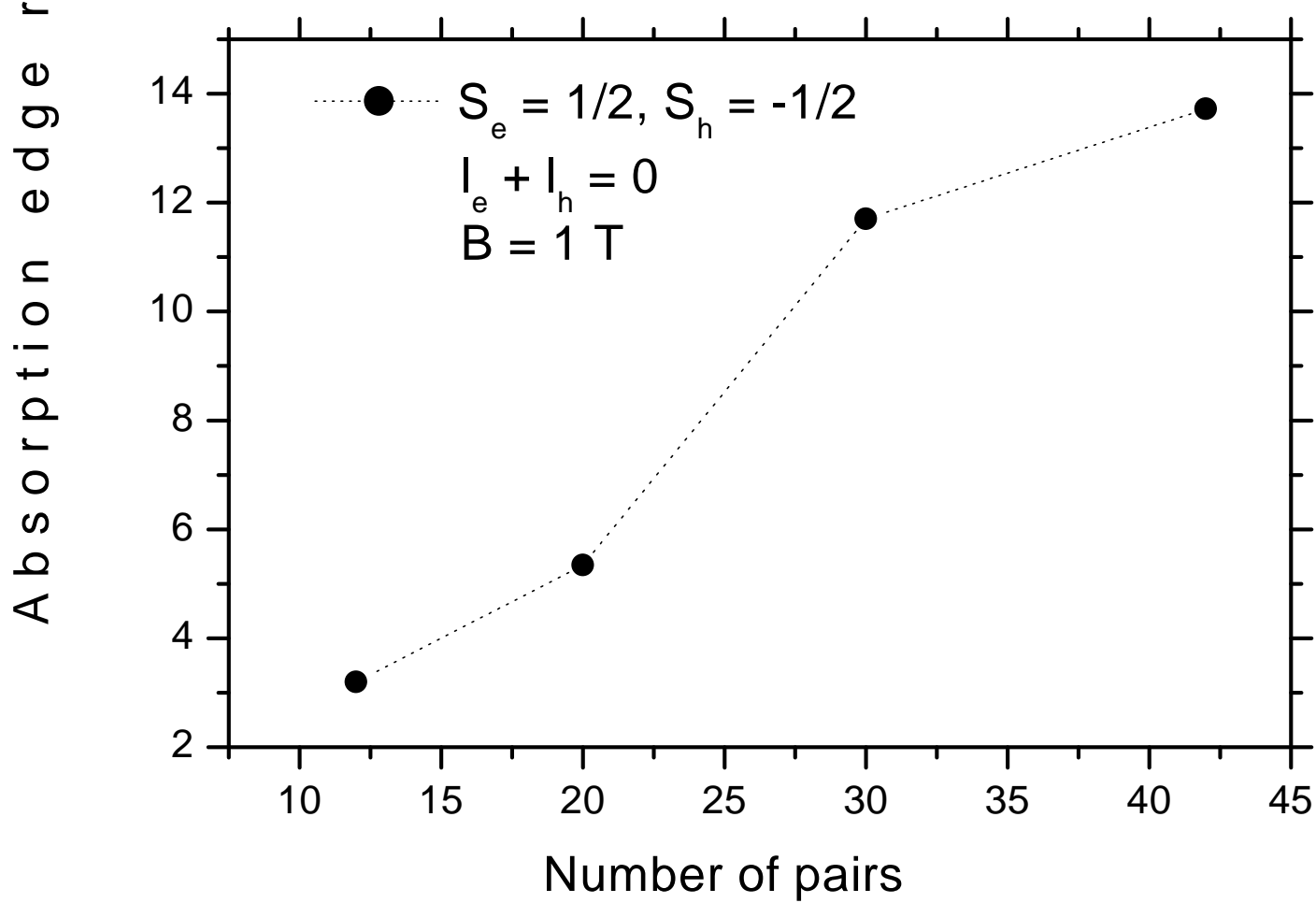
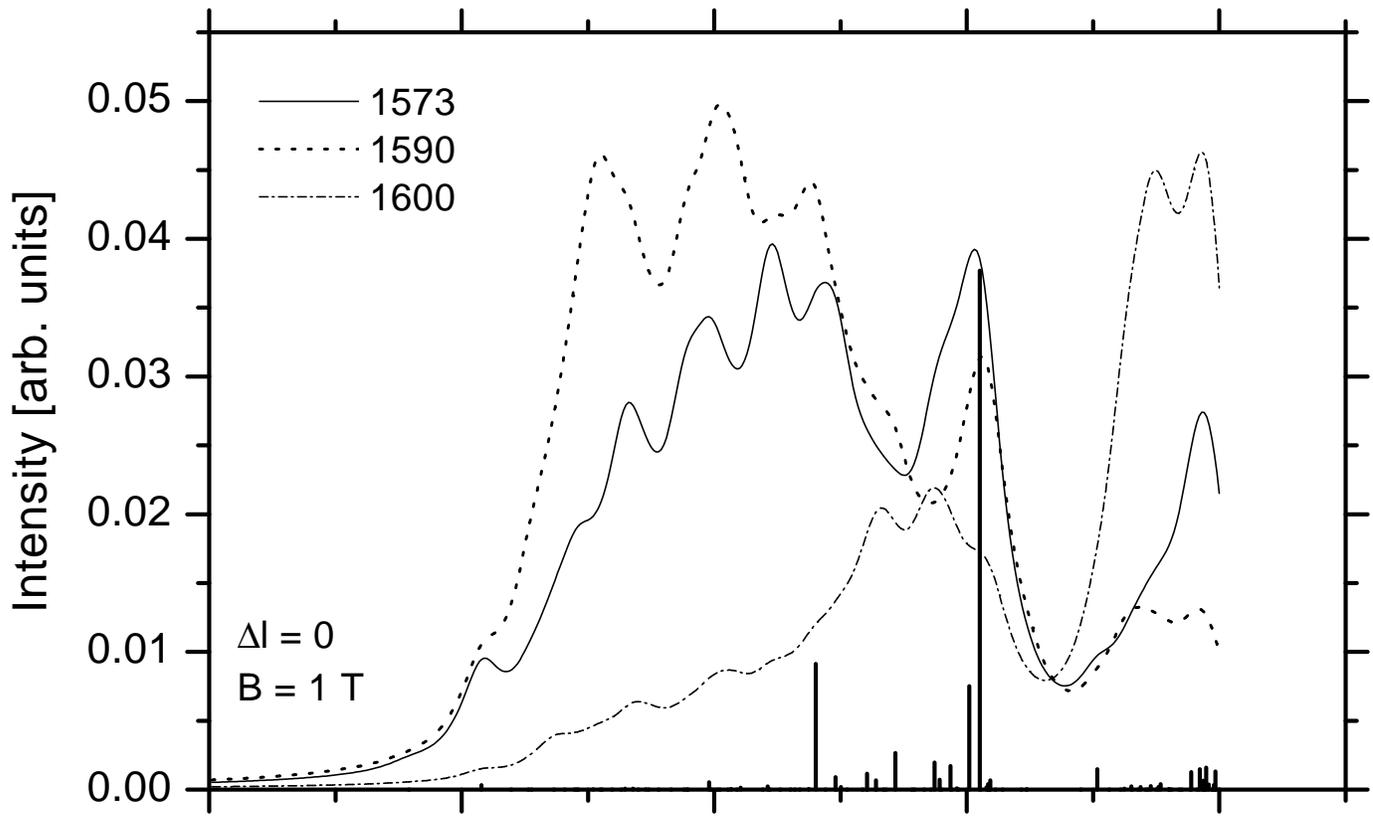
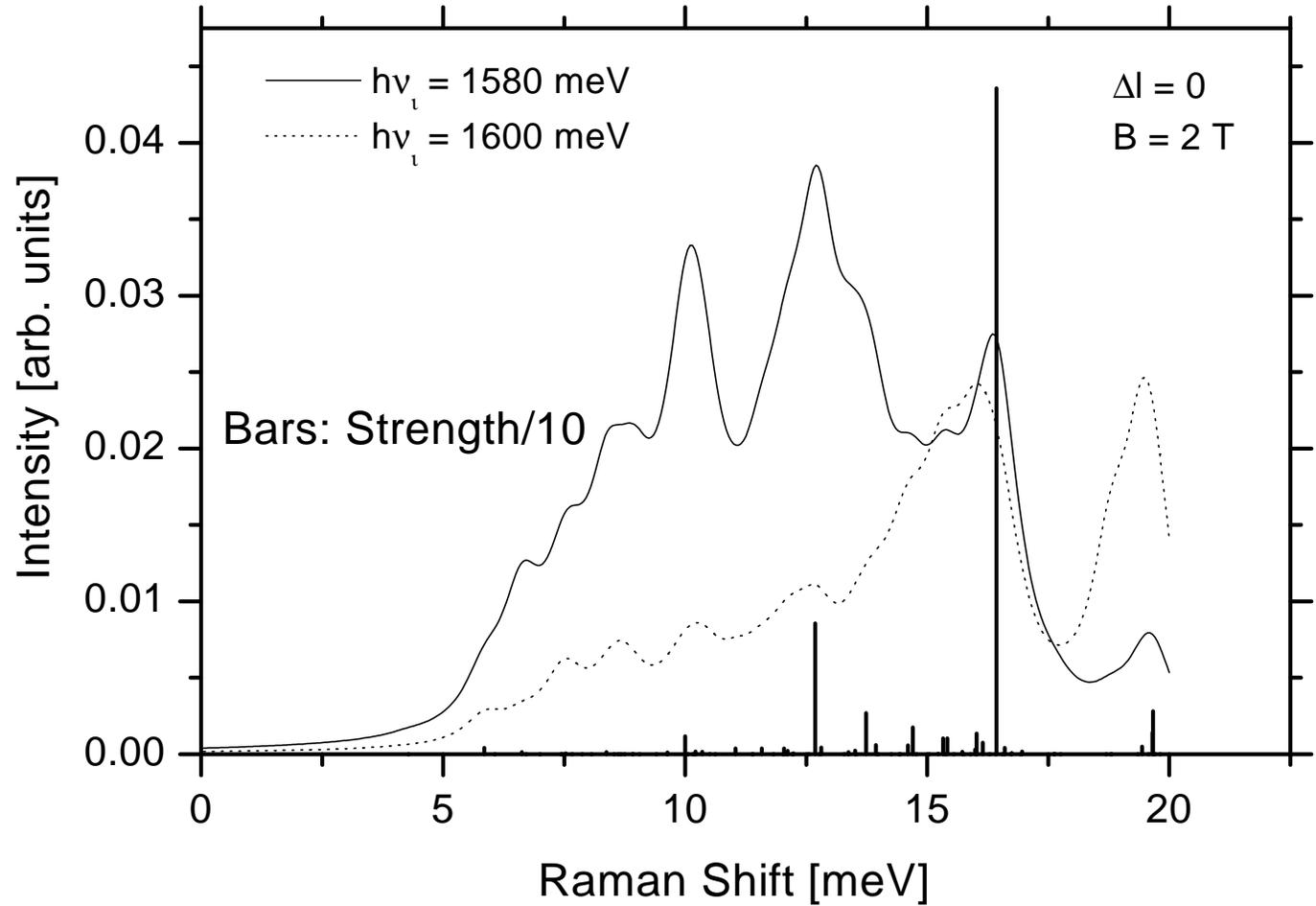
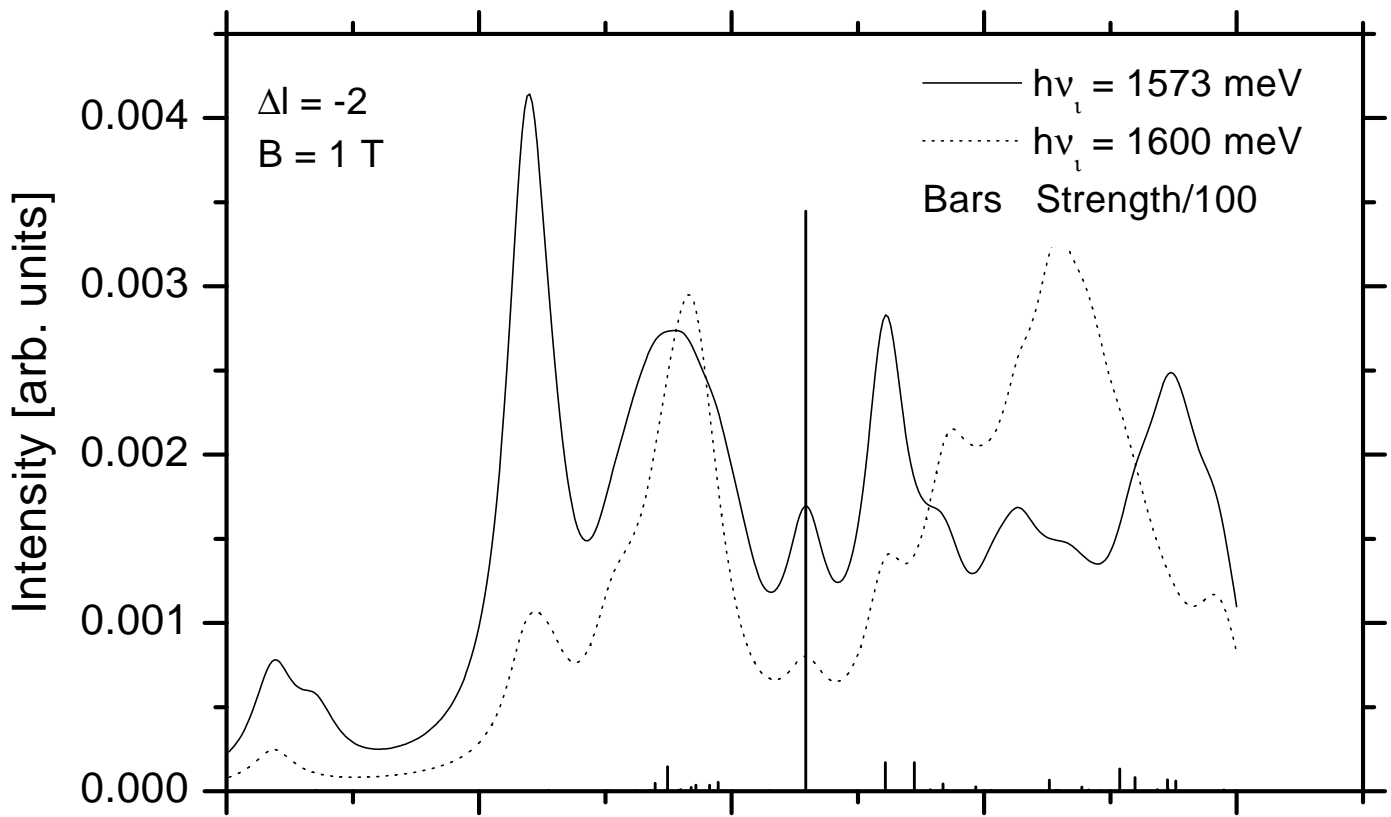
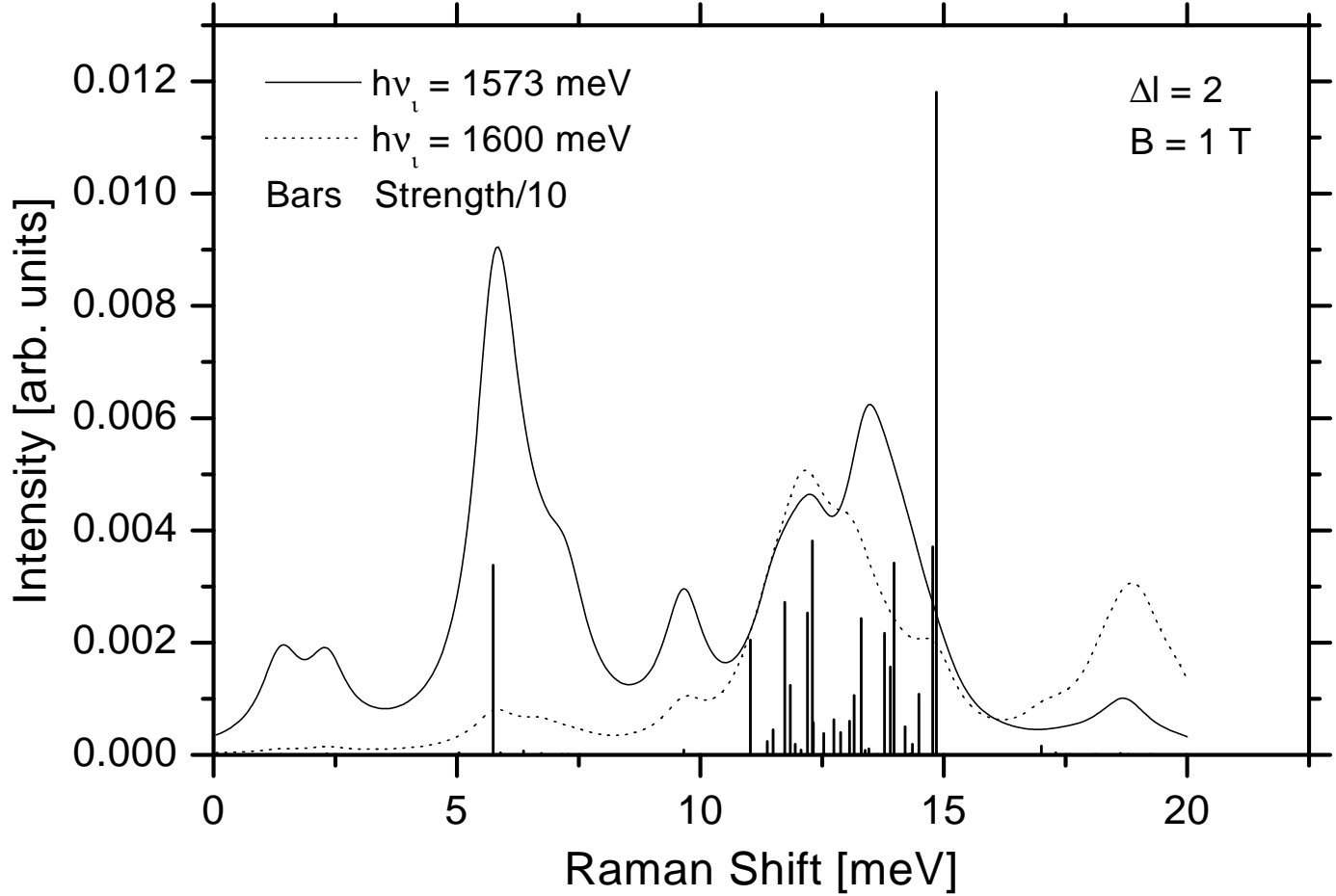
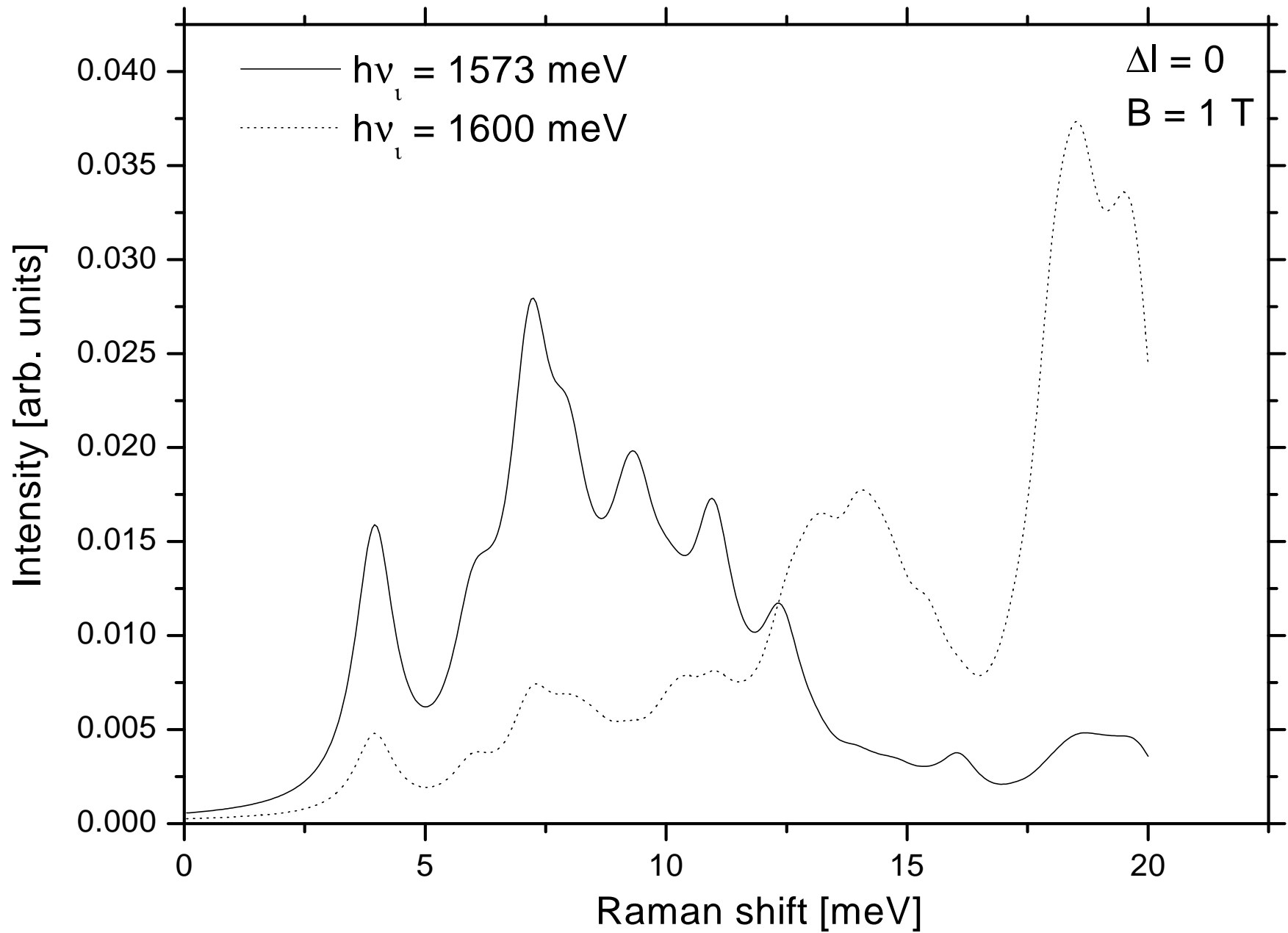
\end{document}